\UseRawInputEncoding
\pdfoutput=1

\documentclass[preprints,article,accept,moreauthors,pdftex]{mdpi} 

\firstpage{1} 
\pubvolume{xx}
\issuenum{1}
\articlenumber{5}
\pubyear{2022}
\copyrightyear{2022}

\history{} 

\usepackage{bookmark}
\usepackage{amsmath} 
\usepackage{amssymb}
\usepackage{graphicx}
\usepackage{float}
\usepackage[normalem]{ulem}
\usepackage{xcolor}
\usepackage{soul}
\setstcolor{red}
\usepackage{cancel}
\usepackage{threeparttable}
\usepackage{subfigure} 
\newcommand\redsout{\bgroup\markoverwith{\textcolor{red}{\rule[0.5ex]{2pt}{1.5pt}}}\ULon}
\Title{Pulsar Glitches: A Review}
\Author{Shiqi Zhou $^{1,5,6}$, Erbil G\"{u}gercino\u{g}lu$^{2}$*, Jianping Yuan $^{3,7}$, Mingyu Ge $^{4}$ and Cong Yu $^{1,5,6}$ }
\AuthorNames{Shiqi Zhou, Erbil G\"{u}gercino\u{g}lu, Jianping Yuan, Mingyu Ge and Cong Yu }
\address{%
$^{1}$ \quad School of Physics and Astronomy, Sun Yat-Sen University, Zhuhai, 519082, China\\
$^{2}$ \quad National Astronomical Observatories, Chinese Academy of Sciences, 20A Datun Road, Chaoyang District, Beijing 100101, China\\
$^{3}$ \quad Xinjiang Astronomical Observatory, Chinese Academy of Sciences, Xinjiang 830011, China\\
$^{4}$ \quad Key Laboratory of Particle Astrophysics, Institute of High Energy Physics, Chinese Academy of Sciences, Beijing 100049, China\\
$^{5}$ \quad State Key Laboratory of Lunar and Planetary Sciences, Macau University of Science and Technology, Macau, China\\
$^{6}$ \quad CSST Science Centre for the Guangdong-Hong Kong-Macau Greater Bay Area, Zhuhai, 519082, China\\
$^{7}$ \quad Centre for Astronomical Mega-Science, Chinese Academy of Sciences, Beijing 100012, China
}
\corres{Email: \url{egugercinoglu@gmail.com} (E. G.)}
\abstract{
$\sim 6\%$ of all known pulsars have been observed to exhibit sudden spin-up events, known as glitches. For more than fifty years, these phenomena have played an important role in helping to understand pulsar (astro)physics. Based on the review  of  pulsar glitches search method, the progress made in observations in recent years is summarized, including the achievements obtained by Chinese telescopes. Glitching pulsars demonstrate great diversity of behaviours, which can be broadly classified into four categories: normal glitches, slow glitches, glitches with delayed spin-ups, and anti-glitches. The main models of glitches that have been proposed  are reviewed and their implications for neutron star structure are critically examined regarding our current understanding. Furthermore, the correlations between glitches and emission changes, which suggest that magnetospheric state-change is linked to the pulsar-intrinsic processes, are also described and discussed in some detail.}
\begin{document}
\section{Searching for Pulsar Glitches}
Pulsars emit radiation across the whole electromagnetic spectrum, and the most plausible geometric explanation for the observed emission is the lighthouse model \cite{Hewish1975}. The electromagnetic radiation from pulsars is originated and beamed from above their magnetic poles. The magnetic axis of a pulsar has a certain angle with respect to its rotation axis. If the earth is located in the range of emission beams of the pulsar, a series of periodic pulse signals are received by telescopes as pulsars rotate. Among them, millisecond pulsars (MSPs) have the greatest long-term stability of period.
The MSP PSR J0437--4715 is considerably more stable than standard atomic clocks \cite{Hartnett2011}. In recent years, the international pulsar timing array project (PTA), which combines observations of a set of pulsars to search for correlated signatures in the arrival times of pulses \cite{Jenet2009}, has achieved unprecedented development. Two best known applications of this project are the detection of ultra-low frequency ($\sim 10^{-9}-10^{-8}\ \rm Hz$) gravitational waves \cite{Hobbs2010} and  establishment of a pulsar-based timescale \cite{Hobbs2020}. Besides, binary millisecond pulsars constitute most extreme physical laboratories and give opportunity to test several theoretical models including the general theory of relativity \cite{bisnovatyi10}.

\begin{figure}[H]
\centering
\includegraphics[width=0.5\textwidth]{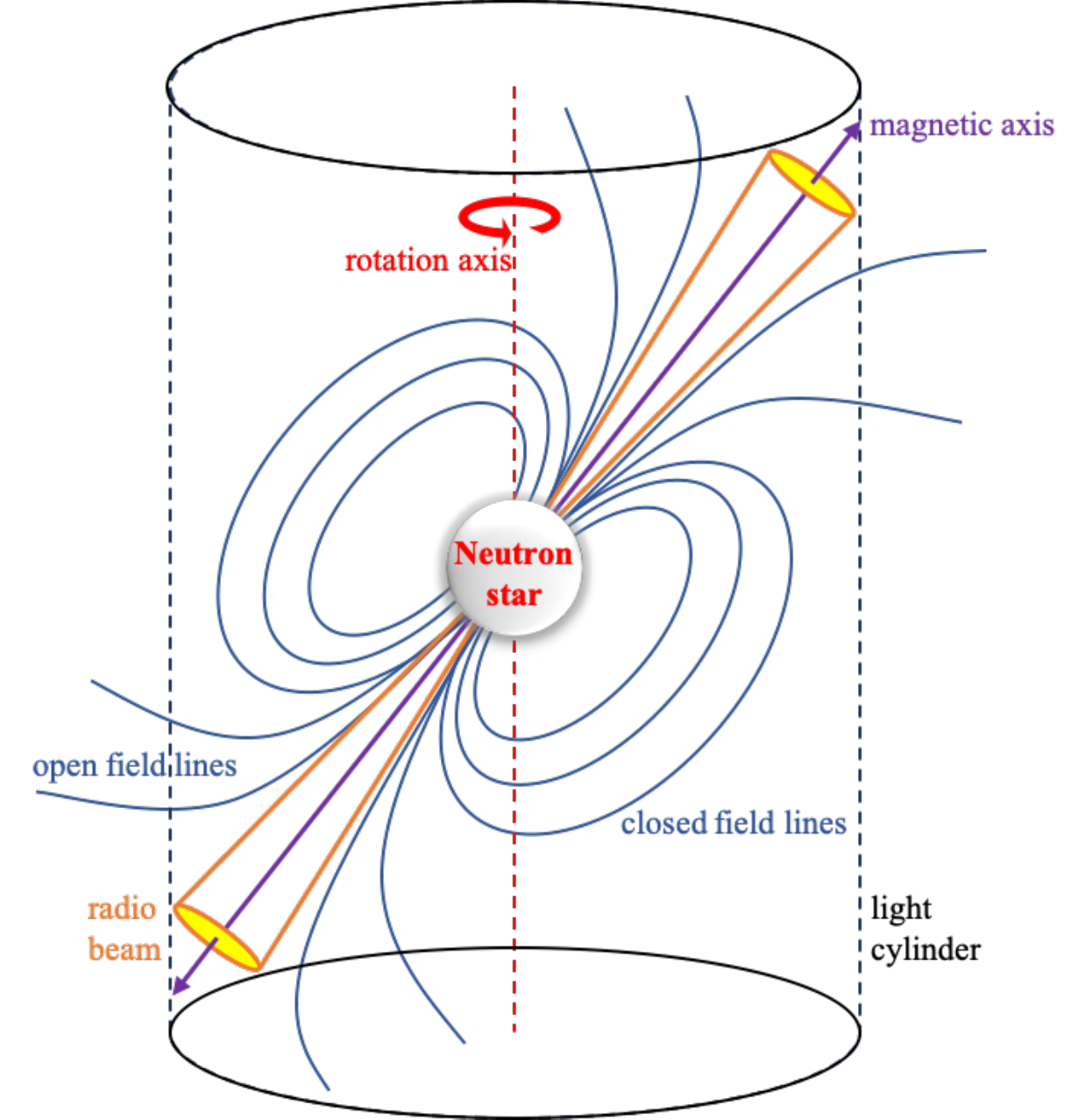}
\caption{The conventional magnetic dipole model: a schematic view of the key features of a pulsar magnetosphere \cite{lorimer2005}.}
\label{emission}
\end{figure}

As of now, there is no fully self-consistent pulsar emission model capable of explaining the structure of the magnetic field and the time-variable phenomena \cite{Harding2018,melrose21,philippov22}. Nevertheless, the simple magnetic dipole model allows a basic understanding of pulsar magnetosphere and the coherent radio emission in pulsars \cite{Gunn1969, Narayan1990, vandenHeuvel2006}. As illustrated in Figure \ref{emission}, an induced electric field exists throughout the magnetised neutron star as a result of its rotation. Under this circumstance, charged particles are lifted out of the solid crust of the star, filling the magnetosphere with a dense plasma \cite{Tchekhovskoy2013}. These particles are accelerated to relativistic velocities, and emit $\gamma$-ray photons by either curvature radiation or inverse Compton scattering on lower energy photons \cite{Ruderman1975, Arons1983, Daugherty1986}. And then, the photons interacting with the magnetic field forms electron-positron pairs \cite{Sturrock1971}. Near the neutron star's polar caps, the flow of particles creates electric currents that produce the observed radio beams. Light cylinder is an imaginary surface at which co-rotation of the closed dipolar magnetic field lines with neutron star breaks-down and field lines open up to space. At light cylinder radius $R_{\rm LC}=cP/2\pi$ with $P$ being the rotational period of the underlying neutron star the plasma speed reaches the speed of light $c$. The currents flowing out from the magnetosphere along the open field lines that cross the light cylinder, as well as the pulsar wind of plasma, exert a torque on the magnetic field lines which then slows the rotation of the neutron star. There is still a lack of reliable observational evidence and theoretical models to accurately describe the coupled magnetic, thermal and spin evolution of pulsars \cite{Baym1969, Goldreich1992, Tauris1998, tong16}. Fortunately, there is a powerful approach known as ''pulsar timing``, which helps to obtain the accurate information on the pulsar spin-down and refine existing radiation models.

%

Pulsar timing is a diagnostic technique in the rotational behaviour by tracking each
pulse's time of arrival (TOA), and involves the evaluation and interpretation of the mismatch between measured and predicted values. Although the individual pulses measured during the observation are highly variable in shape and intensity \cite{Kramer1999}, several hundreds or thousands of pulses can be summed in time, frequency and polarisation to produce a highly stable integrated pulse profile over the timescales of years \cite{Bansal2020}. In order to obtain a series of TOAs, each of the integrated pulse profile is cross-correlated with a high signal-to-noise ratio ``standard'' profile. Usually, rapidly rotating pulsars with narrow pulse profiles have higher accuracy of TOAs. After correcting for ToAs to the Solar System Barycentre in order to remove the Earth's orbital motion effects, the evolution of the phase of the pulse sequence can be expressed as a truncated Taylor series \cite{Edwards2006b}:
\begin{equation}
    \phi(t) = \phi_0 + \nu_0 (t - t_0) + \frac{1}{2!} \dot{\nu}_0 (t - t_0)^2 + \frac{1}{3!} \ddot{\nu}_0 (t - t_0)^3,
    \label{timmod}    
\end{equation}

\noindent where $t$ is the observation time. $\phi_0$, $\nu_0$, $\dot{\nu}_0$ and $\ddot{\nu}_0$ represent the phase, spin-frequency and its first and second derivatives as measured at the reference time $t_0$. Regardless of the effects of matter outflow and accretion, the spin-down of an isolated radio pulsar is mainly dominated by the magnetic dipole radiation, so that the accurate predictions of the arrival time of every one of its pulses can be made from the timing model given by equation (\ref{timmod}).

However in an actual long term observation, it is not always the case. The slow-down law of pulsars predicted by equation (\ref{timmod}) is generally perturbed by two types of timing irregularities: timing noise and glitches \cite{d'alessandro96}. Timing noise is a slow, long-term, discernible, stochastic wandering in the periods of pulsars. In 1969, two independent groups, Radhakrishnan \& Manchester \cite{Radhakrishnan1969} and Reichley \& Downs \cite{Reichley1969} presented the first ever observation of a sudden increase in PSR J0835--4510's (Vela pulsar) spin frequency $\nu$, known as ``glitch''. Obviously, for some pulsars pulse arrival times are not well predicted by the simple spin-down model, reflected in the observed irregularity of the pulsar's rotation. The discovery of glitches have naturally raised the curiosity to search for more similar events using telescopes around the world. Now, high cadence monitoring of an exceedingly large number of pulsars are conducted by the northern and southern hemisphere radio telescopes: 76-m Lovell \cite{basu22} and 64-m Parkes \cite{ljd21}. Most of glitches are detected with these two telescopes. A database of all known pulsar glitches is regularly updated in two different locations: one at ATNF\footnote{\href{https://www.atnf.csiro.au/people/pulsar/psrcat/glitchTbl.html}{https://www.atnf.csiro.au/people/pulsar/psrcat/glitchTbl.html }{$-$ accessed 2022 September}}\label{foot1} \cite{Manchester2005} and also by Jodrell Bank (JBO)\footnote{\href{http://www.jb.man.ac.uk/~pulsar/glitches/gTable.html}{http://www.jb.man.ac.uk/~pulsar/glitches/gTable.html }{$-$ accessed 2022 September}}\label{foot2} \cite{espinoza11}. After combining entries from both the catalogues there are a total of 719 glitches identified in 239 pulsars \cite{Arumugam2022}. As will be seen in Section \ref{Classification} below, these glitches can be further classified into four subclasses of normal glitches, slow glitches, glitches with delayed spin-ups, and anti-glitches.

These observed glitches are modelled using an additive function as below \cite{Edwards2006}:

\begin{equation}
    \label{glitch_pars}
        \begin{aligned}
            \phi_{\rm g} (t) = & \Delta\phi + \Delta \nu_{\rm p} (t - t_{\rm g}) + \frac{1}{2!}\Delta \dot{\nu}_{\rm p} (t - t_{\rm g})^2  \\
            & + \sum_i^N \Delta \nu_\mathrm{d}^{(i)} \tau_\mathrm{d}^{(i)}  \left[ 1 - \exp \left( {-\frac{(t - t_{\rm g})}{\tau_\mathrm{d}^{(i)}}} \right) \right],
        \end{aligned}
\end{equation}
 
\noindent whose value is zero if $t \leq t_{\rm g}$. $\Delta\phi$ is an offset in pulse phase. The positive or negative sign of the permanent jumps in the spin frequency $\Delta\nu_{\rm p}$ at the time of a glitch $t_\mathrm{g}$ represents usual glitches or anti-glitches, respectively. $\Delta\dot{\nu}_{{\rm p}}$ is the permanent change in the frequency derivative (spin-down rate) relative to the pre-glitch value. The fouth term models the $i$ of $N$ exponential recoveries with a transient frequency component $\Delta \nu_\mathrm{d}$ and decay time constant $\tau_\mathrm{d}$. $\Delta \nu_\mathrm{d} > 0$ corresponds to the amplitude of the exponential recovery. $\Delta \nu_\mathrm{d} < 0$ indicates the existence of a delayed spin-up component after the initial unresolved step. As for the criterion of slow glitches, a continuous increase in frequency and sudden decrease followed by a gradual increase in spin-down rate $|\dot{\nu}|$ are recognised from plots of $\nu$ and $\dot{\nu}$ versus time. Hence, the observed absolute jumps $\Delta\nu_{\rm g}$ and $\Delta\dot{\nu}_{\rm g}$ in a glitch are described as:
\begin{equation}
\Delta\nu_{\rm g}=\Delta\nu_{\rm p}+ \sum_i^N \Delta \nu_\mathrm{d}^{(i)},
\end{equation}
\begin{equation}
\Delta\dot{\nu}_{\rm g}=\Delta\dot{\nu}_{\rm p}- \sum_i^N \frac{\Delta \nu_\mathrm{d}^{(i)}}{\tau_\mathrm{d}^{(i)}}.
\end{equation}
In addition, the relative glitch amplitudes in frequency and frequency derivative are $\Delta\nu_{\rm g}/\nu$ and $\Delta\dot{\nu}_{\rm g}/\dot{\nu}$, respectively. The degree of recoveries are quantified by the parameter $Q={\sum_i^N \Delta \nu_\mathrm{d}^{(i)}}/\Delta\nu_{\rm g}$. Noteworthy, the maximum change of the spin frequency $\Delta\nu_{\rm max}$ and $\Delta\dot{\nu}_{\rm max}$ its first time-derivative are derived from comparing the timing solutions away from the spin-up event to describe the slow glitch effect. Uncertainties associated with these parameters are obtained using standard error propagation methods.

\begin{figure}[H]
\centering
\includegraphics[width=0.80\textwidth]{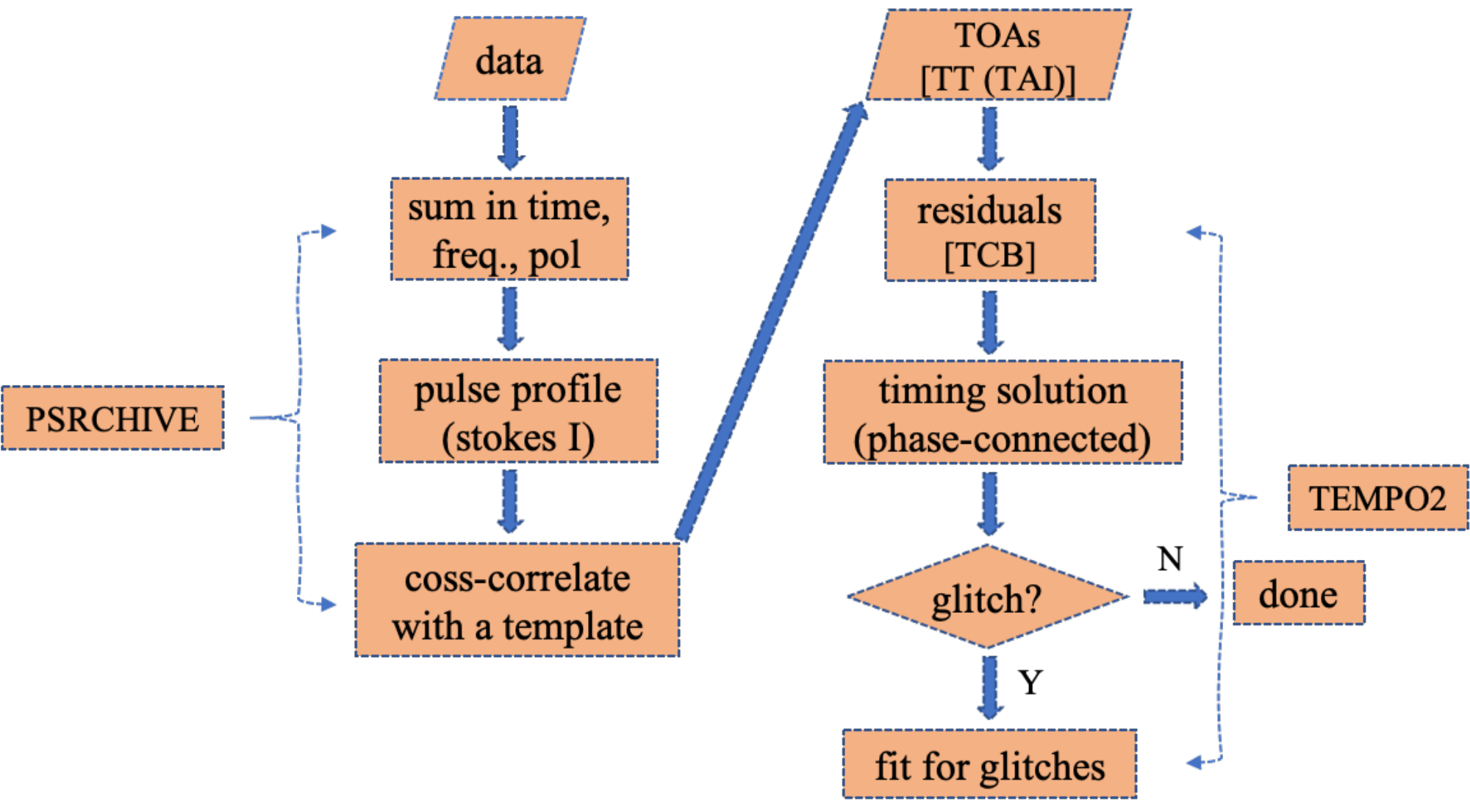}
\caption{The most commonly employed data reduction routine for radio pulsar glitches.}
\label{routine}
\end{figure}

For decades, the usual procedures of data reduction for radio pulsar glitches are as shown in Figure \ref{routine}. The two key steps are the measurements of the TOAs and implementation of phase-connected timing residuals. The popular pulsar data analysis packages \textsc{psrchive} \cite{Hotan2004} provide functionality for generating TOAs; and then, these TOAs are used to fitted with a model that contains a set of known parameters to achieve a phase-coherent timing residuals ($R=(\phi - N)/\nu$, where $N$ is the nearest integer to $\phi$) \cite{Hobbs2006}, using softwares such as \textsc{tempo} \cite{Taylor1979}, \textsc{psrtime} \cite{Hobbs2006}, \textsc{tempo2} \cite{Hobbs2006}, and \textsc{pint} \cite{Luo2021}. \textsc{tempo2} remains by far the most popular software packages in the bunch of pulsar timing tools. Using these tools, the substantial manual efforts are required to obtain a satisfactory phase-connected timing residuals. Until recently, Freire et al. \cite{Freire2018} and Phillips et al. \cite{Phillips2022} have developed new automated algorithms, DRACULA and APT, to determine phase-connected timing solutions for pulsars. Therefore, pulsar glitches can be identified via visual inspection with the sudden phase discontinuity in the timing residuals relative to a solution based on earlier data. Glitches manifest as timing residuals becoming increasingly negative with time and new models are needed for post-glitch behaviours. In the case of frequent occurrence of glitches with particularly large sizes, it is almost impossible to obtain the good post-glitch timing solutions. Melatos et al. \cite{melatos20} presented a complementary method to this problem in the form of the Hidden Markov Model (HMM) algorithm, which measures the permanent changes in spin frequency $\Delta\nu_{\rm p}$ and spin-down frequency $\Delta\dot{\nu}_{\rm p}$ during glitches with a HMM. Dunn et al. \cite{dunn22} employed HMM to analyze low-cadence timing data to constrain the magnitude of missing glitches in pulsars. Singha et al. \cite{Singha2021} presented a real-time automated glitch detection pipeline (AGDP) that enables statistical tests to determine phase coherence between TOAs and has been  implemented at the Ooty Radio Telescope (ORT) to recognise automatically glitches. Soon, the searching  techniques of glitches with massive human intervention will be consigned to the past.

In China, some scientific achievements have been made in the area of glitches research. Table \ref{China} presents the information on the number of pulsars currently timed by Chinese telescopes and the number of glitches discovered by each telescope. Notably, frequent observations spanning 22 years with the Nanshan
26-m radio telescope (NSRT) have detected 103 glitches, which account for about $15\%$ of all published glitches. The first slow glitch of PSR B1822--09 \cite{zww04} and significant changes in pulse profile associated with PSR B2035+36's glitch activity \cite{Kou2018} were discovered by NSRT. Ge et al. \cite{GZL2020} and Zhang et al. \cite{Zhang2018} reported that a delayed spin-up process occurred before the normal recovery process in the November 2017 Crab (PSR B0531+21) glitch with Insight-HXMT (Hard X-ray Modulation Telescope) and XPNAV-1 (X-ray Pulsar Navigation satellite) data, respectively. Currently, about 350 pulsars have long been become a continuously timing target with the largest dish FAST (Five-hundred-meter Aperture Spherical Radio Telescope), which has the capacity to follow more than 2000 pulsars. In the future, next-generation telescopes QTT (Qitai 110-m Radio Telescope) \cite{Cong2017} and JRT (Jingdong 120-m Radio Telescope) \cite{Qian2020} will provide an opportunity to track more pulsars, more frequently, allowing glitches to be investigated in more detail.

\begin{table}[H]
\caption{Chinese telescopes to perform a search for glitch events.}
\begin{center}
\begin{threeparttable}
\begin{tabular}{cccclccl}
\hline
Telescopes     & Location                 & Diameter              & Frequency  
               & Pulsars                  & Start                 & No. of                & Ref.                 \\
               & & & & & & Glitches & \\
\hline
NSRT           & Nanshan                  & 26 m                  & 1.54 GHz            
               & $\sim300$                & 2000                  & 103                           & \cite{zww04, ywm10, Dang2020, Yuan2010, Kou2018, Liu2018, Zhou2022, Yuan2013,Wang2013,Zou2008, Yuan2017, Wang2001, Yuan2016, Wang2012}                              \\
TMRT           & Shanghai                 & 65 m                  & 2.25/4.82/8.60 GHz             
               & $\sim100$                & 2013                  & 5                             & \cite{Liu2019, Liu2021, Liu2022, Liu2020}                     \\
Insight-HXMT   & Space                    & --                    & 1-250 keV             
               & $\sim10$                 & 2017                  & 3                             & \cite{Zhang2022,GZL2020}                     \\
               
KMRT           & Yunnan                   & 40 m                  & 2.25/4.85 GHz            
               & $\sim90$                 & 2008                  & 1                             & \cite{xyl19}                     \\
HRT            & Luonan                   & 40 m                  & 1.40 GHz           
               & $\sim10$                 & 2014                  & 1                             & \cite{Luo2020}                     \\

XPNAV-1        & Space                    & --                    & 0.5-10 keV             
               & $\sim26$                 & 2016                  & 1                             & \cite{Zhang2018}                     \\

   
FAST           & Pingtang                 & 500 m                 & 1.25 GHz 
               & $\sim350$                & 2016                  & --                            & \cite{Han2021}                     \\          
QTT            & Qitai                    & 110 m                 & --            
               & --                       & --                    & --                            & \cite{Cong2017}                     \\
JRT            & Jingdong                 & 120 m                 & --            
               & --                       & --                    & --                            & \cite{Qian2020}                     \\
\hline

\end{tabular}
\end{threeparttable}
\end{center}
\label{China}
\end{table}

\section{Properties of Pulsar Glitches}
Almost all published glitches were identified by visual inspection of the phase residuals, relative to a predicted model. Glitches with fractional sizes of $\Delta\nu/\nu\sim10^{-9}$ do not lose phase coherence over several hundred days, while those with sizes $\Delta\nu/\nu\sim10^{-6}$ may see a change in residuals of a large fraction of the pulse period in just a few days \cite{ymh13}. Once the glitch has been determined, groups of TOAs that overlap are fitted to demonstrate the time evolution of $\nu$ and $\dot{\nu}$ around the glitch. To illustrate the permanent and transient jumps in one glitch, the time-dependence of the frequency residuals $\Delta\nu$ relative to the pre-glitch spin-down model are generally presented as in Panel (a) of Figure \ref{0835}. Also, an expanded plot of frequency residuals $\Delta\nu$ where the mean post-glitch value has been subtracted from the post-glitch data is derived to clearly show the exponentially changing components, as seen in Panel (b). The variations of spin-down rate $\dot{\nu}$ is visualized in detail as shown in Panel (c). Janssen \& Stappers \cite{janssen06} found that the minimum glitch size could be detected with a threshold of $\Delta\nu/\nu\sim10^{-11}$ by performing Monte Carlo simulations to test the glitch detection method on PSR J0358+5413 \cite{Millhouse2022}. The millisecond pulsar J0613--0200 has suffered a glitch with a size $\Delta\nu/\nu\sim2.5\times10^{-12}$, which is the smallest that has been recorded \cite{mjs16}. Several studies have indicated that the previously published pulsar timing results did not include all glitches in individual pulsar or in particular datasets \cite{janssen06, yu17, Jankowski2019}. The following reasons may result in incomplete observations of glitches:
\begin{itemize}
\item[1.] Given the limited resources available to observatories and the large numbers of pulsars, timing observations of some pulsars are not carried out around the glitch \cite{SKL2021}.
\item[2.] Making low-cadence, short dwell time observations near a $\dot{\nu}$ transition epoch may result in misinterpreting  the timing behaviour, leading one to omit a glitch. \cite{Shaw2018b, Dang2020}.
\item[3.] Glitches are too small to be resolved due to being below the present limit of detectability \cite{espinoza11}.
\item[4.] Pulsars that exhibit high levels of timing noise may have small glitches that go undetected in the data \cite{ywm10}.
\item[5.] If small glitches occur during the initial stages of recovery following a large glitch, they could be missed because of the need of timing fit \cite{WBL2001,espinoza14}.
\item[6.] When a glitch is identified to occur in a significant observation gap, it is impossible to distinguish between a single glitch and multiple, smaller glitches \cite{yu17,janssen06}.
\end{itemize}

\begin{figure}[H]
\centering
\includegraphics[width=0.6\textwidth]{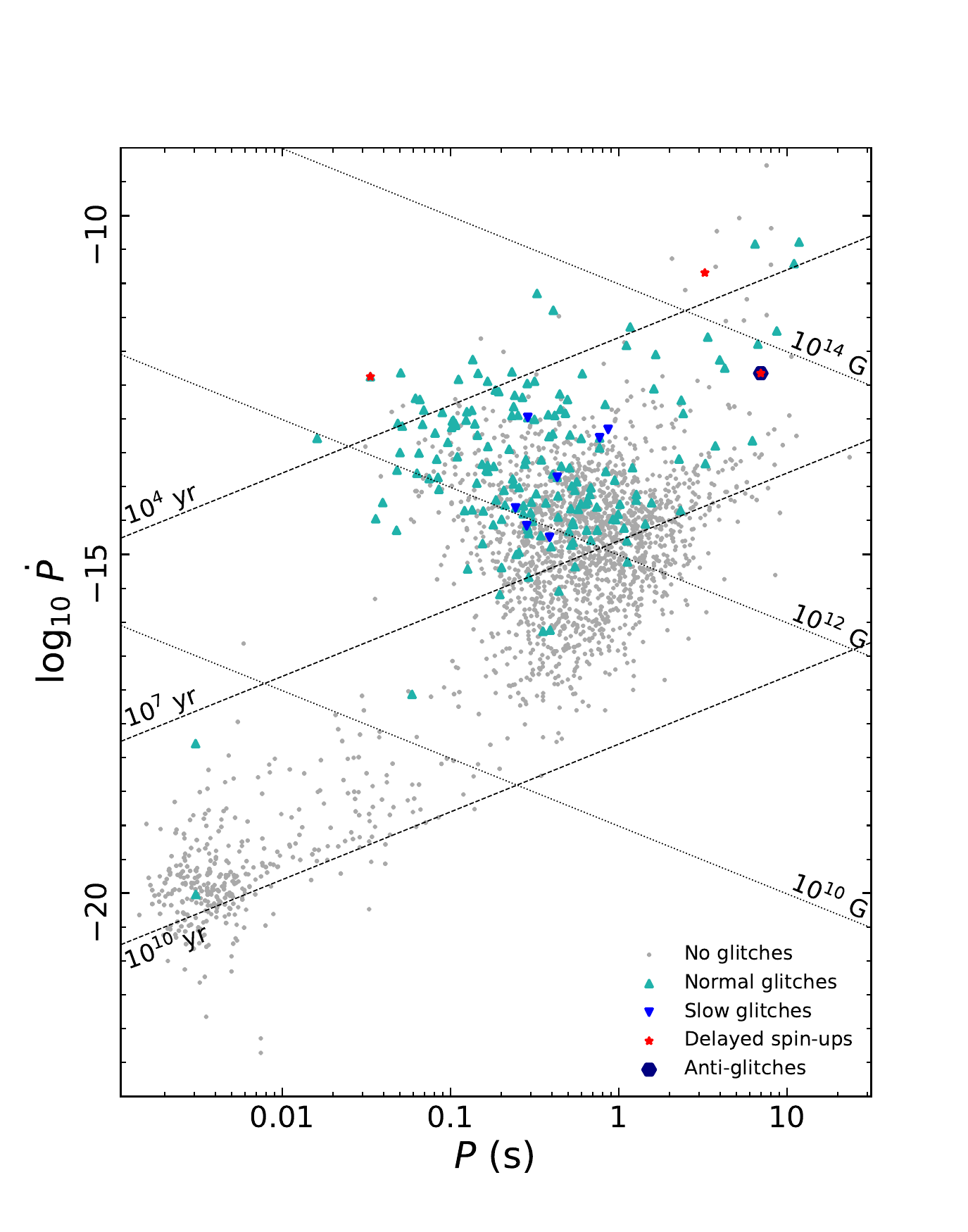}
\caption{Period–period-derivative ($P-\dot{P}$) diagram showing the pulsars for which glitch is detected based on the ATNF Pulsar Catalogue \cite{Manchester2005}. The various sub-classes of glitching pulsars are represented by the markers in the legend.}
\label{ppdot}
\end{figure}

Around $6\%$ of known pulsars have been observed to glitch at least once since the first discovery of glitch in 1969. Out of the 239 glitching pulsars, 144 have only glitched once or twice. The known glitching pulsars are shown on the the period–period-derivative ($P-\dot{P}$) diagram in Figure \ref{ppdot}. Clearly, glitches occur predominantly in young pulsars (characteristic ages $\sim10^{3}-10^{7}\,\rm yr$ \cite{ymh13}) located in the top right of the diagram, but the size of and time between glitches do not depend only upon the pulsar’s position in the $P-\dot{P}$ diagram \cite{Hobbs2002}. The curious glitches phenomena have been discovered from ordinary  pulsars, to super-powerful magnetic field harbouring magnetars \cite{yrb20}, to millisecond pulsars \cite{Cognard2004}, and are predominantly seen from young and high spin-down rate pulsars. PSRs J0534+2200 (Crab), J0537--6910, J0835--4510 (Vela), J1341--6220, J1740--3015 are remarkable examples of displaying frequent glitches. 
A significant linear correlation between glitch size $\Delta\nu$ and time to next glitch has been established for the big glitcher PSR J0537--6910 \cite{antonopoulou18, ferdman18}, which has undergone 53 glitches by September 2022 \cite{ho20}. The Vela and Crab pulsars are the most widely studied glitching pulsars. The Vela pulsar, first observed in 1968, has since then had 24 glitches, with a frequency jump of $\Delta\nu/\nu \sim 10^{-6}$ for most

\noindent of them. The post-glitch decay starts out exponential relaxation, with different timescales from 1 minute to several hundred days, but it eventually exhibits linear recovery of spin-down rate in time with a constant second time derivative of spin frequency \cite{DLM2007, ywm10}. In contrast, the glitch sizes of the Crab pulsar ($\Delta\nu/\nu \sim 10^{-7}-10^{-9}$) are about at least an order of magnitude smaller than those for the Vela pulsar. There is usually a sudden, exponential decay in the post-glitch behavior over several days, as well as a persistent increase in $\dot{\nu}$ \cite{Lyne1993,ywm10}. Long-term timing observations of Crab- and Vela-like pulsars revealed their glitch behaviours were similar to the two pulsars. More interestingly, two micro-glitches have been observed from the millisecond pulsars (MSPs J0613--0200 \cite{mjs16} and B1821--24 \cite{Cognard2004}), which are thought to be the most stable type of pulsar population. Even so, if sufficient post-glitch timing observations are available to model the glitches, the influence on the timing precision for pulsar timing array applications, such as gravitational wave (GW) detection, can be eliminated. McKee et al. \cite{mjs16} indicated that the rate of glitches for MSPs is significantly lower than that of the general population, with a probability of $\sim 50 \%$ that another glitch will be seen in a timing array programme of pulsar within 10 years. Moreover, glitches occurred in magnetars are well known to have large sizes, and sometimes coincide with emission changes and outbursts \cite{dib14}. 31 magnetars have been dicovered, 7 of which have exhibited a total of 22 glitches\footnote{\href{http://www.physics.mcgill.ca/~pulsar/magnetar/main.html}{http://www.physics.mcgill.ca/~pulsar/magnetar/main.html }{$-$ accessed 2022 September}} \cite{Olausen2014, Jawor2022}. Chukwude \& Urama \cite{chukwude10} found that glitches have amplitudes $\Delta\nu/\nu\gtrsim10^{-9}$ and associated decreases in $\dot\nu$, while micro-glitches are smaller in amplitude ($\Delta\nu/\nu\lesssim10^{-10}$) and can have either positive or negative changes in $\dot\nu$. 
According to Yu et al. \cite{ymh13}, after a glitch, part of the sudden change in both $\nu$ and $\dot{\nu}$ often recovers exponentially, and $\dot{\nu}$ then continues to decay linearly until the next glitch.

\section{Classification of Glitches}\label{Classification}

\subsection{Normal Glitches}
\begin{figure}[H]
\centering
\includegraphics[width=0.8\textwidth]{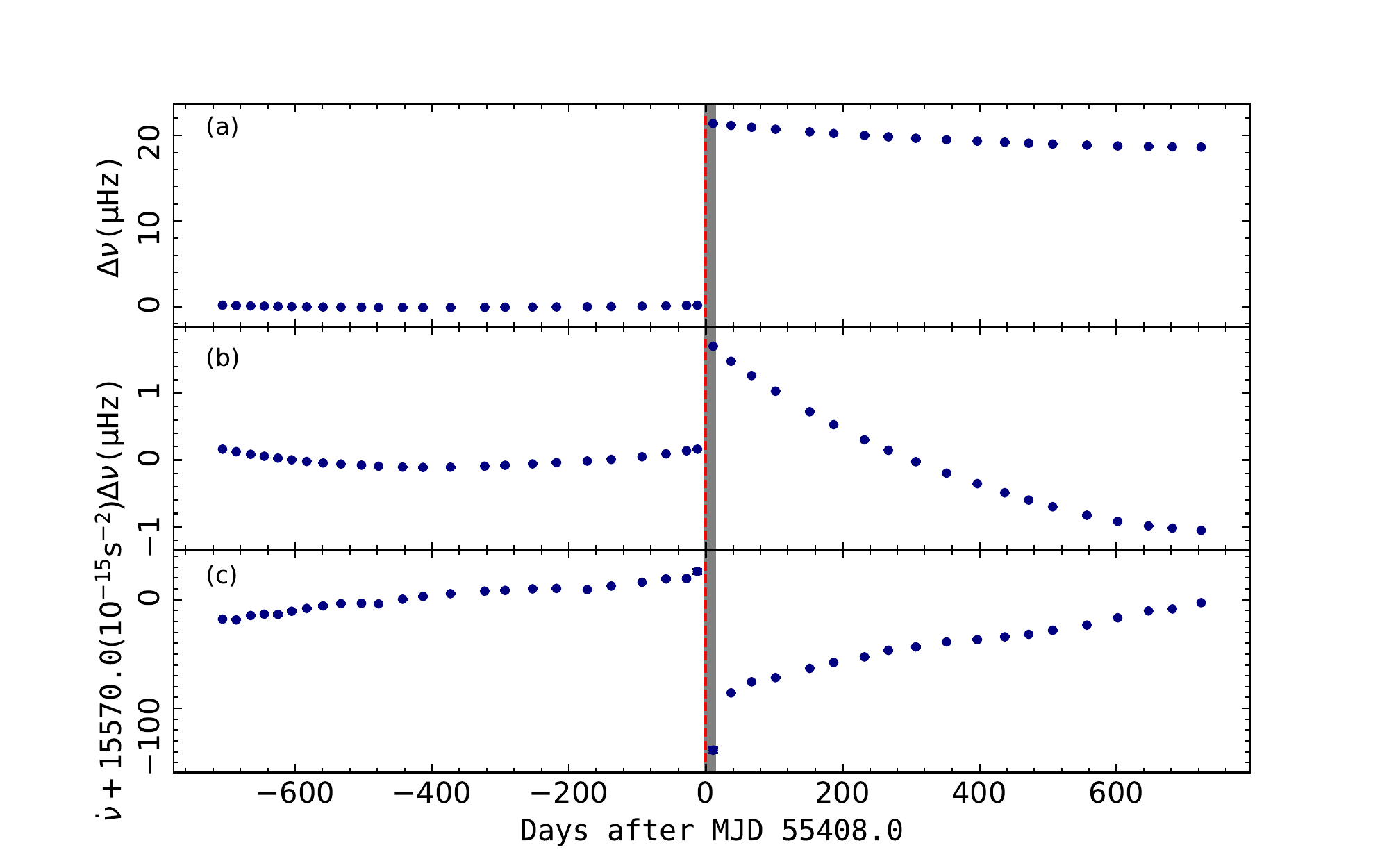}
\caption{The PSR J0835--4510's (Vela pulsar) July 2010 glitch with an exponential decay timescale of $13(2)\rm \ d$ (marked grey area) observed by the Fermi-LAT \cite{egy22, ljd21}: (a) the variations in spin frequency after subtracting the pre-glitch timing model; (b) expanded plot of $\Delta\nu$ where the mean post-glitch value has been subtracted from the post-glitch data; (c) the evolution of $\dot{\nu}$. The vertical line indicates this glitch epoch at MJD 55408(1).}
\label{0835}
\end{figure}

To-date, more than 200 pulsars have collectively shown hundreds of (> 650) normal glitches. The fractional amplitude $\Delta\nu/\nu$ of these events are in the range of $10^{-12}$ \cite{mjs16} to $10^{-3}$ \cite{sss17}. Near-instantaneous rise times and lengthy recoveries are common features in these glitches. Figure \ref{0835} shows a typical glitch with an exponential decay timescale of $13(2)\rm \ d$ in Vela puslar (PSR J0835–4510) occurred in July 2010. The rapid spin-up timescales of glitches can be assumed to be less than $5\rm \ s$ \cite{PDH2018}, $30\rm \ s$ \cite{dlm07}, and $40\rm \ s$ \cite{dml02}, according to the results of glitch analysis for the Vela pulsar carried out by the southern hemisphere observers. Often, the instantaneous rise is followed by a temporary increase in the spin-down rate. For the post-glitch behaviours, the response of the spin frequency is diverse.
Across the population of glitches, the evolution of the post-glitch rotation rates is usually modelled as a combination of exponential and linear recoveries to the pre-glitch values \cite{hhm16, ymh13} . The former last from a few days to several months and the latter persist on a time-scale of years. Perhaps a more striking aspect of these glitches is that the impermanent part has been observed to require multiple decaying exponentials for the best timing fit. PSRs J0835--4510 (ten times) ,  J1119--6127 (once), J1757--2421 (once), J1803--2137 (once) and J2337+6151 (once) have shown evidence of two exponential decay terms in same glitch \cite{Urama2002, ymw17, xyl19, Zubieta2022}. More specifically, Flanagan \cite{F90} reported that the December 1988 Vela glitch best explained with three exponential terms and Dodson et al. \cite{dml02} confirmed four exponential terms occurred in the January 2000 Vela glitch. In the early days following the first glitch observation in the Vela pulsar Baym et al. \cite{baym69} proposed the ``two component model'' that normal glitches are the outcome of the relaxation of the neutron-star superfluid after crust quakes to the current equilibrium oblateness. Glitch recoveries are thought to be first evidence for the existence of a superfluid interior of the neutron star \cite{bf11, baym69}. Soon after Anderson \& Itoh \cite{anderson75} suggested that normal glitches are driven by the transfer of angular momentum from unpinned vortex lines within superfluid to the solid crust. Arguably, normal glitches contain invaluable information on the interior of neutron stars. Therefore, general glitch behaviours can be used to investigate not only the structure of pulsar magnetosphere, but also the dynamical process operating inside a neutron star.

\subsection{Slow Glitches}

\begin{figure}[H]
\centering
\includegraphics[width=0.8\textwidth]{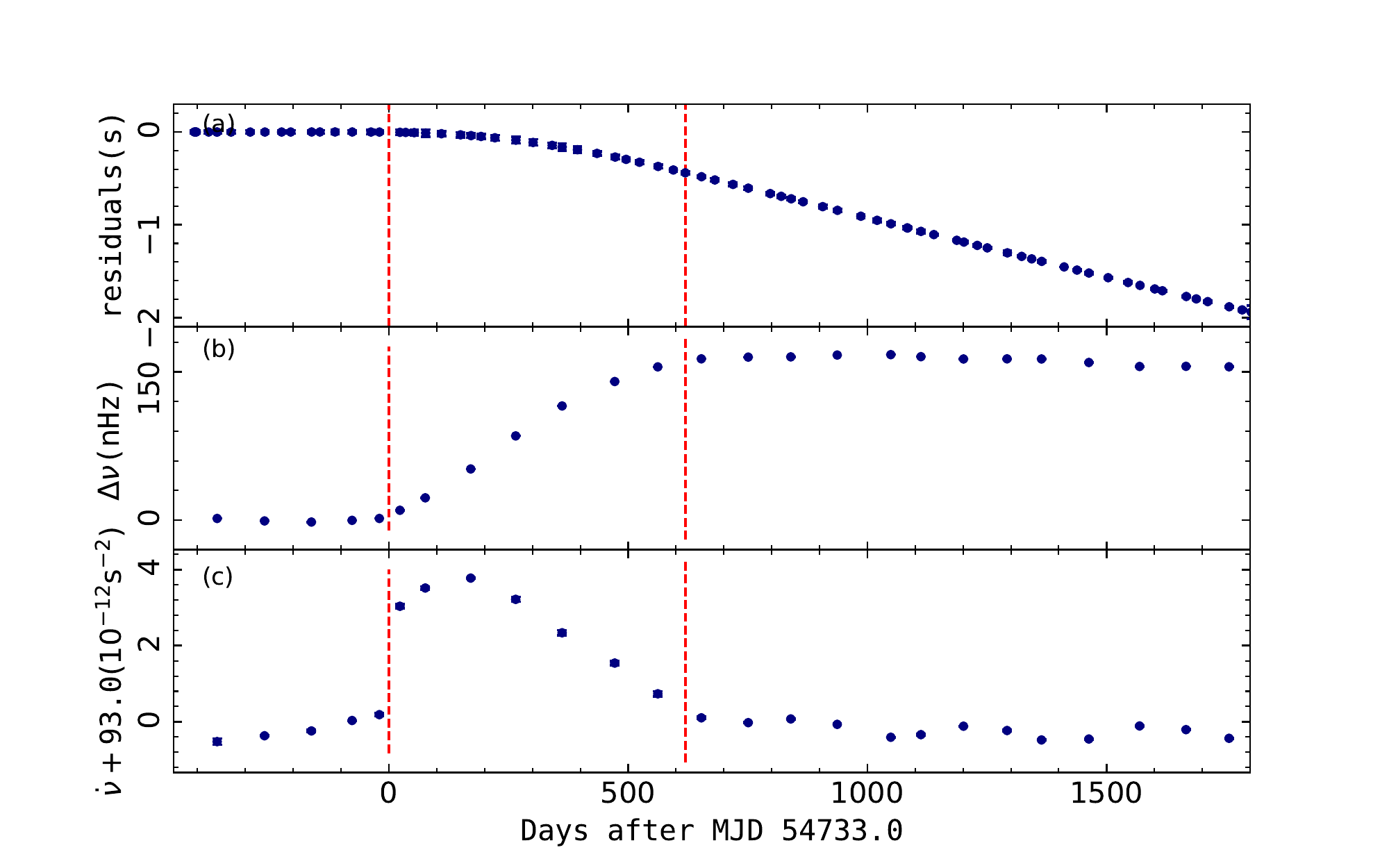}
\caption{A very large slow glitch in PSR J1062--5100 \cite{zzz19}: (a) timing residiuals with respect to the pre-glitch spin-down solution; (b) variations of the frequency residual $\Delta\nu$ after subtracting the pre-glitch spin-down model. (c) observed variations of $\dot{\nu}$. The two red dashed vertical lines indicate that the slow glitch occurred between MJD 54733 and 55330.}
\label{slow}
\end{figure}

A total of 31 slow glitches have been observed in 7 pulsars \cite{zzz19,sjb22}, for which the spin frequency $\nu$ slowly increases whereas the spin-down rate $\dot\nu$ decreases over several months before \textcolor{red}{$\dot\nu$} relaxes back to its original state over a slightly longer time scale (Figure \ref{slow}) \cite{ywm10,zww04,s98}. This process repeats periodically, and the peak value of the glitch amplitude is the same each time. The maximum changes in the spin frequency $\nu$ and spin-down rate $|\dot{\nu}|$ for these slow glitches range from $2.3-46 \rm \ nHz$ and $0.15\times10^{-15}-3.15\times10^{-15} \rm \ s^{-2}$ ; the corresponding relative maximum changes are $\Delta\nu/\nu \sim 0.9\times10^{-9}-31.2\times10^{-9}$ and $|\Delta\dot{\nu}|/|\dot{\nu}|\sim 1.8\times10^{-3}-23.6\times10^{-3}$. The occurrence of slow glitches and normal glitches in the PSR B1822--09 is of special interest \cite{s09}. Zhou et al. \cite{zzz19} linked the variations of the integrated pulse profile to the detection of a very large slow glitch in PSR J1602--5100. At the beginning and end of this slow glitch there was 
the presence and absence of a new trailing component. Theoretically, the angular momentum exchange model \cite{acp89,eyc17}, which is currently widely accepted to explain the physical mechanism of normal glitches, cannot be applied to the process of slow events. Link \& Epstein \cite{link96} believed that the slow glitches are the result of a sudden increase in the temperature of the inner crust of the neutron star. However, Hobbs et al. \cite{hlk04} suggested that the slow glitches are just a manifestation of timing noise, which is significantly dominated by sustained random wandering in either the phase, spin, or spin-down rate \cite{bgh72}. That means they cannot be classified as a new subclass of pulsar glitches, and may be more widespread throughout the pulsar population. However, slow glitches have not been detected in any other glitching pulsars so far. While curious, the lengthy rise time, periodic recurrence and uniform amplitude of these slow glitches strongly indicate that they are a separate, though possibly related, category of event.

\subsection{Delayed Spin-ups}

\begin{figure}[H]
\centering
\includegraphics[width=0.8\textwidth]{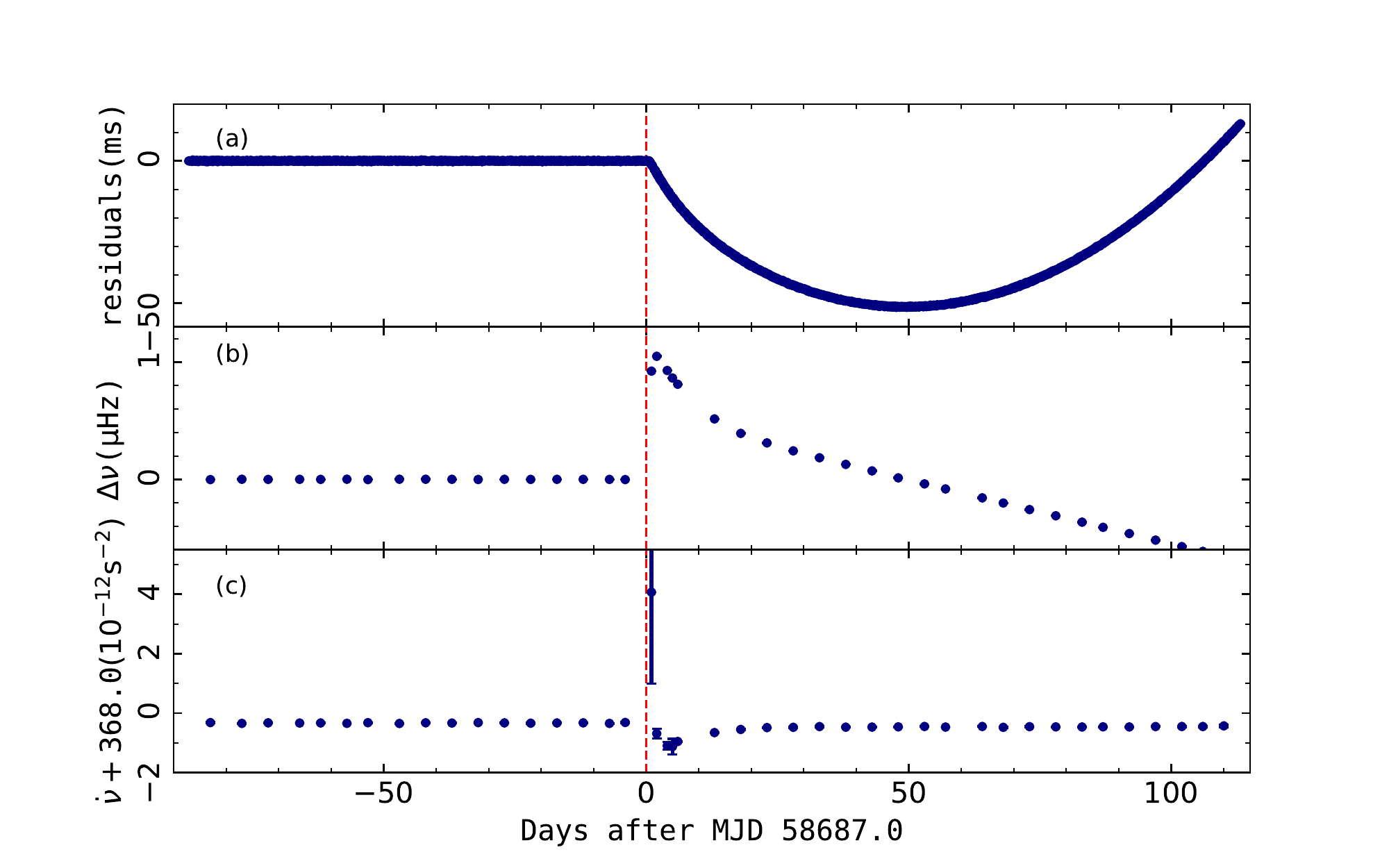}
\caption{The simulated July 2019 Crab delayed spin-up behaviour as presented by Shaw et al. \cite{SKL2021}: (a) timing residiuals relative to a pre-glitch spin-down model; (b) the time-dependence of the frequency residuals $\Delta\nu$; (c) the evolution of spin-down rate $\dot{\nu}$. The vertical line signifies the glitch around MJD 58687.565(4).}
\label{delayed}
\end{figure}

Figure \ref{delayed} demonstrates the evolution of the rotation frequency $\nu$ and its first derivative
$\dot\nu$ in the delayed spin-up of the 2019 Crab pulsar glitch. It is obvious that glitch with delayed spin-up component can be recognized by a short period of the $\nu$ increase exponentially immediately after the glitch occurrence and followed by rapid decay of $\dot\nu$. At present, delayed spin-up processes have been detected in the glitches of three pulsars, Crab (six times) \citep{LSP1992, WBL2001, SLS2018, GZL2020, SKL2021}, 1E 2259+586 (once) \citep{WKT2004} and SGR J1935+2154 (once) \citep{GZL22}. The timescale of these fast rising transient components ranges between 0.5 days and 14.1 days. The capture of delayed spin-up component in the multiple glitches from the Crab pulsar are partly attributed to the high and regular observing cadence at the Jodrell Bank Observatory. On the contrary, if post-glitch transient variation are prevalent in glitches, the low and irregular cadence observations are mainly why they are not detected. Although the upper limit of the rising time scale for the Vela pulsar glitches have been constrained to within 5 s \cite{PDH2018} by southern hemisphere observers monitoring near-continuously, no delayed spin-up components have been observed. This may be another evidence that the glitches of Crab and Vela pulsars have a different origin \cite{M2018}.
It is worth noting that these observed delayed spin-up behaviours were only detected in young pulsars during their big glitches, and there is a slow exponential recovery process after the delayed spin-up. In particular, the total 2020 SGR J1935+2154 glitch magnitude including the delayed increase is the largest among the all glitches observed to date, and the transient decay timescale $\sim 8\rm \ d$ is much longer than that of the Crab's events. Ge et al. \cite{GZL22} found that the pulse profile of SGR J1935+2154 changed rapidly and drastically after the unusual large glitch. Combined with the detection of repeated FRB 200428 after the glitch with a delay of likely 3 days, Ge et al. \cite{GZL22} argue that internal changes in young magnetars generate large glitch to significantly alter the structure of the magnetosphere and trigger X-ray bursts and FRB 200428. Just recently Younes et al. \cite{younes22} reported a possible connection between a new glitch occurred on  October 5, 2020 and three FRB-like radio bursts in SGR J1935+2154.
There is a power-law ($\alpha = 2.0$) correlation between $\Delta\nu/\nu$ and the extended spin-up timescale $\tau$ of 6 glitches in 3 pulsars suggesting that the mechanisms behind these unusual glitches are similar \cite{GZL22}. The finding is also interpreted as generally consistent with the G{\"u}gercino{\v{g}}lu \& Alpar \cite{erbil19} theory. Moreover, it is impossible to tell whether delayed spin-ups are only related to large jumps, because if the delay time scale is too short or small jumps are too small to be observed by existing observation equipment. In short, the detection of delayed spin-ups offers a unique opportunity to study the micro-physical processes governing interactions between the neutron star interior and the crust. It was also proposed that SGR J1935+2154 glitch may arise from the ejection of a particle wind emanating from close to magnetar polar cap which effectively extracts angular momentum from the star \cite{younes22}.

\subsection{Anti-glitches}

The so-called anti-glitches are first coined when Archibald et al. \cite{akn13} observed a sudden negative jump of spin frequency in the magnetar IE 2259+586, accompanied by a X-ray outburst and a significant spin-down rate change. Recently, Ray et al. \cite{rgh19} reported three large anti-glitches in the accreting ultraluminous X-ray source NGC 300 ULX1 during its most recent outburst. Ducci et al. \cite{Ducci2015} evaluated the possibility that anti-glitches occur in accreting pulsars, and found that these spin-down glitches may take place more often in the superfluid vortex scenario. Until now, only a special class of pulsars, magnetars exhibiting contemporaneous radiative changes, have been observed to undergo anti-glitches. In total 7 anti-glitches are detected in 3 pulsars \cite{akn13,sinem14,yrb20,ibs12,rgh19,Pintore2016,Vurgun2019}. An \& Archibald \cite{An2019} speculated a possible anti-glitch occurred in magnetar CXOU J164710.2–455216 from the interrupted phase of residuals. Undoubtedly, the unexpected occurrence of anti-glitches are strongly challenging the standard glitch theories. The close correlation observed between spin-down glitches and outburst activity suggest that these events are related to magnetospheric activity \cite{M2018}. Tong \cite{Tong2014} used the wind braking model to successfully interpret the anti-glitch behaviours of magnetar 1E 2259+586. Huang \& Geng \cite{Huang2014} suggested that anti-glitches occur as a result of the collision of a small mass body with pulsars. Garcia \& Ranea-Sandoval \cite{Garcia2015} interpreted that anti-glitches is an obvious consequence of the cumulative decay of the internal toroidal magnetic field component resulting in the rearrangement of the stellar structure. Howitt \& Melatos \cite{howitt22} performed simulations in vortex avalanches scenario to find that anti-glitches are caused by interrupting the secular increase of the angular velocity. Whatever the cause, rethinking of the mechanism behind the glitches of all pulsars is needed. 
\begin{figure}[H]
\centering
\includegraphics[width=0.8\textwidth]{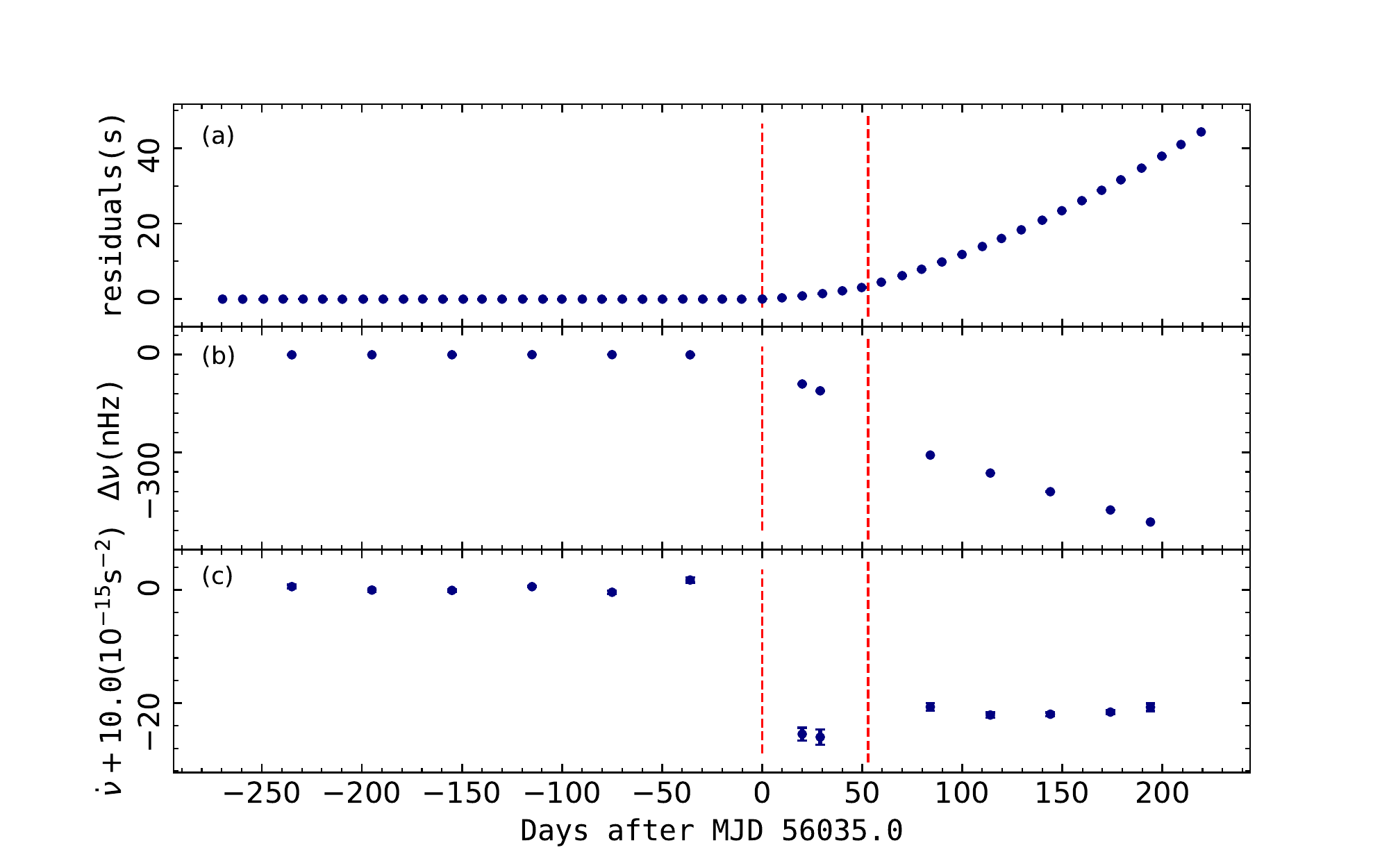}
\caption{The simulated two anti-glitch in magnetar IE 2259+586 around 2012 \cite{akn13} : (a) timing residiuals relative to a spin-down model before anti-glitch; (b) the frequency residuals $\Delta\nu$; (c) the spin-down rate $\dot{\nu}$ of before and after anti-glitch. The red vertical dashed lines mark the glitch epochs: 56035 \cite{akn13}, 56088.4 \cite{hph14}.}
\label{anti}
\end{figure}
\section{Models of Pulsar Glitches}\label{models}
The properties of rotational glitches in a given pulsar, especially the relaxation of the spin frequency to a value slightly less than its extrapolated original pre-glitch level sheds light into the multicomponent structure of neutron stars and leads to the view that glitches represent the exchange of angular momentum between these components. The literature on the mechanisms responsible for pulsar glitches is immense. Immediately following the first detection, pulsar glitches have been attributed to a wide variety of mechanisms by invoking processes occurring both in the magnetosphere and internal to the neutron star, such as accretion, magnetospheric instabilities, crustquakes, corequakes and catastrophic unpinning of vortex lines \cite{pines74}. It was also conjectured that pulsar glitches may result from neutronization in the envelopes of neutron stars following loss of angular momentum and shallow increase in matter density \cite{bisnovatyi70}. Another possibility for large scale vortex unpinning is the vortex avalanches wherein an unpinned vortex segment encounters a pinned vortex and causing it to unpin which unpins other pinned vortices due to superfluid flow and collective unpinning develops in this way. Such a scenario is observed in simulations albeit for low number of vortices and point defects, i.e. lattice nuclei, compared to a real neutron star case \cite{warszawski08,warszawski11,warszawski12,warszawski13,khomenko18,lonnborn19,howitt20}. In the early days of pulsar glitch research several experiments were devised with superfluid Helium II filled containers accelerated and decelerated occasionally by hand to mimic neutron star rotational evolution \cite{tsakadze72,tsakadze75,tsakadze79,tsakadze80}. In recent years numerical simulations have been conducted to visualize several aspects of pulsar glitch recovery  \cite{larson02,sidery10,vaneysden11,vaneysden12,xie13,vaneysden13,vaneysden14,howitt15,graber17}.    

\subsection{Basics of Superfluid Vortex Dynamics for Neutron Star Rotational Evolution}

Superfluidity is at the heart of the most glitch models \cite{anderson75,alpar84a,jones93,sedrakian95,link96,ruderman98,sedrakian99,pizzochero11}. Moreover, pulsar timing noise may arise from erratic coupling of superfluid with the normal matter inside a neutron star \cite{alpar86,jones90d,haskell11,link14}. We begin this section by reviewing some key concepts of superfluid vortex lines for neutron star rotational dynamics and glitches. Those readers who want to learn more about the superfluidity of neutron stars can refer to the general review articles in the literature \cite{baym79,shaham80,sauls89,pines99,haskell15,haskell18,sedrakian19}.

A superfluid can maintain rotation only if it is perforated with vortex lines. The number, distribution and interaction of these vortex lines fully determine the rotational behaviour of superfluid inside a container. In the case of neutron stars the container is its solid crust. 

The equation of the observed crust's rotation rate is given by the torque equilibrium on the neutron star due to decelerating external braking torque $N_{\rm ext}$ and internal superfluid torque $N_{\rm int}$ as

\begin{equation}
I_{\rm n}\dot\Omega_{\rm n}=N_{\rm ext}+N_{\rm int}=N_{\rm ext}-I_{\rm cs}\dot\Omega_{\rm cs}-I_{\rm core}\dot\Omega_{\rm core},
\label{creos}
\end{equation}
where the subscripts "n", "cs" and "core" refer to normal matter, crustal superfluid, and core superfluid respectively, and $I$ and $\dot\Omega$ are the moments of inertia and spin-down rates of the corresponding components. Recall that the rotation rate and spin frequency are related to each other via $\Omega=2\pi\nu$. In order to find the spin-down rate of the superfluid components one has to consider the superfluid's circulation due to velocity field around each vortex line:
\begin{equation}
    \oint \vec v_{\rm s}.d\vec\ell=2\pi r^{2}\Omega_{\rm cs}=N_{\rm v}\kappa=\int\int n_{\rm v}\kappa dS=\int_{0}^{r}2\pi r'n_{\rm v}(r')\kappa dr'.
    \label{vorcir}
\end{equation}
Here $\kappa=h/2m_{\rm n}$ with $h$ and $m_{\rm n}$ being Planck constant and bare neutron mass, respectively is the vorticity quantum attached to each line and $N_{\rm v}$ is the total number of vortex line inside a closed contour. If one takes spatial derivative of equation (\ref{vorcir}) and considers rigid rotation for superfluid, then 
\begin{equation}
\frac{\partial\left(2\pi r^{2}\Omega_{\rm cs}\right)}{\partial r} =4\pi r\Omega_{\rm cs}=2\pi r n_{\rm v}\kappa\rightarrow n_{\rm v}=\frac{2\Omega_{\rm cs}}{\kappa}.
\label{vorden}
\end{equation}
Equation (\ref{vorden}) implies that areal density of vortex lines are fully determined by the rotation rate of the superfluid and in order to superfluid to spin down the vortices should be expelled from a given surface area. The time evolution of the vortex number density is given by the continuity equation:
\begin{equation}
    \frac{\partial n_{\rm v}}{\partial t}+\vec\nabla.\left(n_{\rm v}\vec v_{\rm L}\right)=0.
\end{equation}
If one takes the time derivative of equation (\ref{vorcir}) and uses equation (\ref{vorden}) then 
\begin{align}
    \frac{\partial}{\partial t}\left(\oint \vec v_{\rm s}.d\vec\ell\right)&=2\pi r^{2}\dot\Omega_{\rm cs}=\frac{\partial}{\partial t}\left(\int\int n_{\rm v}\kappa dS\right)=-\int\int [\vec\nabla.(n_{\rm v}\kappa v_{\rm L})]dS \nonumber\\&=-\int\frac{1}{r}\frac{d}{dr}\left(r n_{\rm v}\kappa v_{\rm L,r}\right)2\pi r dr,
\label{vortime}
\end{align}
where $v_{\rm L,r}$ is the vortex line's velocity component in the radial direction. Equation (\ref{vortime}) leads to spin-down law for a superfluid:
\begin{equation}
    \dot\Omega_{\rm cs}=-\frac{n_{\rm v}\kappa v_{\rm L,r}}{r}=-\frac{2\Omega_{\rm cs}}{r}v_{\rm L,r},
    \label{sfsd}
\end{equation}
where in the last step we have used equation (\ref{vorden}) for the vortex line number density. Equation (\ref{sfsd}) tells us that in order for superfluid to keep up with the rotation rate of the normal matter crust and thus spin down its vortices should migrate radially outward. Therefore, the rate of the superfluid spin change is fully dependent on the $v_{\rm L,r}$. In order to find the radial vortex line migration rate we consider the Magnus force equation,
\begin{equation}
    \vec{F}=\rho_{\rm s}\vec{\kappa}\times\left(\vec{v_{\rm s}}-\vec{v_{\rm L}}\right),
    \label{mageq}
\end{equation}
together with the force due to magnetised vortex electron scattering of the form \cite{alpar84b,alpar88,sedrakian95a}
\begin{equation}
\vec{F}=C\left(\vec{v_{\rm c}}-\vec{v_{\rm L}}\right),
\label{corefir}
\end{equation}
The solution of system of equations (\ref{mageq}) and (\ref{corefir}) gives
\begin{equation}
    v_{\rm L,r}=\frac{\left(v_{\rm s}-v_{\rm c}\right)}{\left[\left(\frac{\rho_{\rm s}\kappa}{C}\right)\right]+\left[\left(\frac{C}{\rho_{\rm s}\kappa}\right)\right]}.
    \label{corelr}
\end{equation}
From equations (\ref{sfsd}) and (\ref{corelr}) the core superfluid obeys the spin-down law
\begin{equation}
    \dot\Omega_{\rm core}=-\frac{2\Omega_{\rm core}}{r}v_{\rm L,r}=-2\Omega_{\rm core}\frac{\left(\Omega_{\rm core}-\Omega_{\rm c}\right)}{\left[\left(\frac{\rho_{\rm s}\kappa}{C}\right)\right]+\left[\left(\frac{C}{\rho_{\rm s}\kappa}\right)\right]}\equiv-\frac{\left(\Omega_{\rm core}-\Omega_{\rm c}\right)}{\tau_{\rm core}},
\end{equation}
where the core superfluid's relaxation time is defined as
\begin{equation}
    \tau_{\rm core}=\frac{\left[\left(\frac{\rho_{\rm s}\kappa}{C}\right)\right]+\left[\left(\frac{C}{\rho_{\rm s}\kappa}\right)\right]}{2\Omega_{\rm core}}.
    \label{taucore}
\end{equation}
For the above considered process of magnetised vortex scattering of electrons $\tau_{\rm core}\simeq(10-200)P$ seconds \cite{sidery09}. The observed fast relaxation of the overshooting post-glitch spin rate after the 2016 Vela glitch within a minute \cite{ashton19} implies time variable coupling of the core superfluid to the observed crustal rotation rate \cite{erbil20}.
The form of radial motion of vortex lines in the crustal superfluid depends crucially on their microscopic interaction with the lattice nuclei \cite{alpar84a,link93}. In the vortex creep model \cite{alpar84a,alpar89}, $v_{\rm L,r}$ is given by a microscopic trial rate of vortex motion around lattice nuclei times the transition probability due to finite temperature of the crust proportional to the Boltzmann factors, i.e.
\begin{equation}
  v_{\rm L,r}=v_{0}\left[\exp{\left(-\frac{E_{\rm p,out}}{kT}\right)}-\exp{\left(-\frac{E_{\rm p,in}}{kT}\right)}\right],
  \label{lrcreep}
\end{equation}
where $v_{0}\sim10^{5}-10^{7}$\,cm s$^{-1}$ is the microscopic trial velocity of a vortex line around nuclear clusters throughout the neutron star crust \cite{erbil16} and, $E_{\rm p,out}$ and $E_{\rm p,in}$ are defined as
\begin{equation}
 E_{\rm p,out}=E_{\rm p}-\Delta E= E_{\rm p}\left(1-\frac{\omega}{\omega_{\rm cr}}\right),
 \label{Epout}
\end{equation}
and
\begin{equation}
 E_{\rm p,in}=E_{\rm p}+\Delta E= E_{\rm p}\left(1+\frac{\omega}{\omega_{\rm cr}}\right).
 \label{Epin}
\end{equation}
Here $E_{\rm p,in}$ is the pinning energy for vortex-nucleus interaction and $\omega_{\rm cr}$ is the maximum velocity difference between the superfluid and the normal matter that a vortex line can withstand before it unpins. Note that when the crustal temperature drops below a few $10^{5}$\,K the vortex line motion against potential barrier sustained by nuclear clusters proceeds through quantum tunnelling rather than classical thermal creep \cite{link93}. 
From equations (\ref{lrcreep}), (\ref{Epout}) and (\ref{Epin}) we find radial component of the vortex line in the crustal superfluid as
\begin{equation}
  v_{\rm L,r}=2v_{0}\exp{\left(-\frac{E_{\rm p}}{kT}\right)}\sinh{\left(\frac{E_{\rm p}}{kT}\frac{\omega}{\omega_{\rm cr}}\right)}.
  \label{vcreep}
\end{equation}
From equations (\ref{sfsd}) and (\ref{vcreep}) the crustal superfluid spin-down rate becomes
\begin{equation}
    \dot\Omega_{\rm cs}=-\frac{4\Omega_{\rm cs}}{r}v_{0}\exp{\left(-\frac{E_{\rm p}}{kT}\right)}\sinh{\left(\frac{E_{\rm p}}{kT}\frac{\omega}{\omega_{\rm cr}}\right)}.
    \label{dotsfcreep}
\end{equation}

The angular velocity difference $\omega\equiv\Omega_{\rm cs}-\Omega_{\rm c}$ between the rotation rates of the crustal superfluid $\Omega_{\rm cs}$ and the observed crust $\Omega_{\rm c}$ evolves as \cite{alpar89,erbil17}
\begin{equation}
    \dot\omega=\frac{I}{I_{\rm c}}\left[-\frac{N_{\rm ext}(t)}{I}-\frac{4\Omega_{\rm cs}}{r}v_{0}\exp\left(-\frac{E_{\rm p}}{kT}\right)\sinh\left(\frac{E_{\rm p}}{kT}\frac{\omega}{\omega_{\rm cr}}\right)\right],
\end{equation}
from equations (\ref{creos}) and (\ref{dotsfcreep}). 

The differential rotation between the normal and the superfluid components dissipates some of the rotational energy of a neutron star and heats up its interior. The heating rate is determined by the difference between the work done by the external braking torque on the neutron star and the change in the rotational energies of the superfluid and normal components as \cite{alpar84a,shibazaki89,alpar98,link99} \begin{equation}
\dot{E}(t)=N_{\rm ext}\Omega_{\rm c}(t)-\frac{d}{dt}\left[\frac{1}{2}I_{\rm c}\Omega_{\rm c}^{2}(t)+\frac{1}{2}I_{\rm cs}\Omega_{\rm cs}^{2}(r,t)\right]=I_{\rm cs}|\dot\Omega_{\infty}|\omega(r,t),
\end{equation} 
where in the last step we used the fact that in rotational equilibrium all the components are decelerating at the same rate, i.e. $\dot\Omega_{\rm c}=\dot\Omega_{\rm cs}\cong\dot\Omega_{\infty}$.

\subsection{Vortex Creep Model}
Vortex pinning becomes favourable if the energy cost per particle inside vortex normal core is smaller when the vortex overlaps with the lattice nucleus than the line stays outside it. Early calculations indicate that in the density range $3\times10^{13}$ to $2\times10^{14}$\,g\,cm$^{-3}$ vortex pinning is energetically favourable \cite{alpar77}. The dragging of the superconducting protons by the superfluid neutrons in the neutron star core generates a proton supercurrent circulating around each vortex line which makes vortex lines strongly magnetized \cite{alpar84b}. This leads effective scattering of electrons from vortices and thus maintains tight coupling of the neutron star liquid core to the solid crust. Hence, the only component inside the neutron star that is weakly coupled to the crust and may be responsible for the long-term relaxation after the glitches is the inner crust superfluid in which vortex lines can get pinned to lattice nuclei.    

From equation (\ref{dotsfcreep}) we can define a linearity parameter as
\begin{equation}
    \eta\equiv\frac{r|\dot\Omega_{\infty}|}{4\Omega_{\rm cs}v_{0}}\exp{\left(\frac{E_{\rm p}}{kT}\right)}.
\label{linnl}
\end{equation}
For $\eta\leq 1$ crustal superfluid is said to be in the linear regime and since $\sinh{x}\cong x$ we obtain
\begin{equation}
\dot\Omega_{\rm cs}=-\frac{E_{\rm p}}{kT}\frac{4\Omega_{\rm cs}}{r\omega_{\rm cr}}v_{0}\exp{\left(-\frac{E_{\rm p}}{kT}\right)}\omega\equiv\frac{\omega}{\tau_{\rm l}},
\end{equation}
where the linear creep relaxation time-scale $\tau_{\rm lin}$ is defined as
\begin{equation}
\tau_{\rm lin}=\frac{r\varpi}{4\Omega_{\rm cs}v_{0}}\exp{\left(\frac{E_{\rm p}}{kT}\right)}.
\label{taulin}
\end{equation}
In the last step $\varpi$ is shorthand notation for
\begin{equation}
    \varpi\equiv\frac{kT}{E_{\rm p}}\omega_{\rm cr}.
\end{equation}

In the linear creep regime the angular velocity lag assumes the form
\begin{equation}
    \omega_{\rm lin}(t)=\exp\left(-\frac{t}{\tau_{\rm lin}}\right)\left[\omega(0)-\frac{1}{I_{\rm c}}\int_{0}^{t}\exp\left(\frac{t'}{\tau_{\rm lin}}\right)N_{\rm ext}(t')dt'\right],
\end{equation}
with $\omega(0)$ being the initial value of the lag at some time $t=0$, while the response of the observed crustal rotation rate to $\omega(0)$ is
\begin{equation}
    \dot\Omega_{\rm c}(t)=\frac{N_{\rm ext}(t)}{I_{\rm c}}+\frac{I_{\rm cs}}{I\tau_{\rm lin}}\exp\left(-\frac{t}{\tau_{\rm lin}}\right)\left[\omega(0)-\frac{1}{I_{\rm c}}\int_{0}^{t}\exp\left(\frac{t'}{\tau_{\rm lin}}\right)N_{\rm ext}(t')dt'\right].
\label{sdrlin}
\end{equation}
Due to the existence of coupling between the external pulsar braking and internal superfluid torques equation (\ref{sdrlin}) may give rise to over-relaxation, i.e. the spin-down rate after the exponential recoveries are over is greater than the pre-glitch original value $Q\gtrsim1$ \cite{erbil17}, which accounts for the observed case seen after many magnetar glitches displaying radiative changes \cite{dib14}.  

In the case of constant external braking torque $N_{\rm ext}=I|\dot\Omega_{\infty}|$, $\omega_{\rm lin}$ simplifies to the steady state, time independent equilibrium value
\begin{equation}
[\omega_{\infty}]_{\rm lin}=|\dot\Omega_{\infty}|\tau_{\rm lin},
\end{equation}
and the post glitch spin-down rate given by equation (\ref{sdrlin}) now becomes
\begin{equation}
    \dot\Omega_{\rm c}=\dot\Omega_{\infty}-\frac{I_{\rm cs}}{I}\frac{\delta\omega}{\tau_{\rm lin}}\exp\left(-\frac{t}{\tau_{\rm lin}}\right),
\end{equation}
where $\delta\omega$ is the change of the lag at the time of the glitch and is given by
\begin{equation}
    \delta\omega=\omega(0)-\omega_{\infty}.
\end{equation}

Equation (\ref{taulin}) is related to the exponential decay time-scale of the transient post-glitch increase in the spin-down rate of the observed crust. Thus, post-glitch observations of pulsars can be used to constrain microphysical properties of the neutron star crust and superfluid traits \cite{alpar89}. Note that the response of vortex pinning and creep across flux tubes to a spin up glitch also leads to exponential decay \cite{erbil14,chamel20}. The exponential decay timescales of solitary pulsars are found to be in qualitative agreement with the response of neutron star core whereas the case of magnetars implies that direct Urca process mediated fast cooling may operate inside this class of objects \cite{erbil17a}. Exponential decay time-scales for magnetars estimated by the model fit observations very nicely if the internal core temperature of these class of neutron stars are below the temperatures predicted by the standard cooling calculations \cite{yakovlev04,page06}. Accelerated cooling maintained by the direct Urca process gives right post-glitch recovery time-scales. Since the observed post-glitch exponential decay time-scales are at most of order $\tau_{\rm lin}\approx100$ d. \cite{ymh13,ljd21,lzz21}, the steady state lag before a glitch is $[\omega_{\infty}]_{\rm lin}\ll\omega_{\rm cr}$ for reasonable threshold values for vortex unpinning $\omega_{\rm cr}\cong10^{-2}-10^{-1}$ rad\,s$^{-1}$ \cite{alpar84a,seveso16} in these linear creep regions. Therefore, linear creep regions cannot be responsible for the glitches by initiating collective vortex unpinning avalanche \cite{alpar89}.  

At the other extreme if $\eta$ given by equation (\ref{linnl}) is greater than one, vortex creep is in the nonlinear regime. For the nonlinear creep stage since $\sinh{x}\approx\exp(x)/2$ equation (\ref{dotsfcreep}) becomes
\begin{equation}
    \dot\Omega_{\rm cs}=-\frac{2\Omega_{\rm cs}v_{0}}{r}\exp\left(\frac{\omega}{\varpi}-1\right),
\end{equation}
and the time evolution of the lag is now given by
\begin{equation}
    \dot\omega=-\frac{I\varpi}{2I_{\rm c}\tau_{\rm lin}}\exp\left(\frac{\omega}{\varpi}\right)-\frac{N_{\rm ext}(t)}{I_{\rm c}}.
\end{equation}
The steady state lag in the nonlinear regime is given by
\begin{equation}
 [\omega_{\infty}]_{\rm nl}=\omega_{\rm cr}\left[1-\left(\frac{kT}{E_{\rm p}}\right)\ln\left(\frac{4v_{0}\tau_{\rm sd}}{r}\right)\right]\simeq\omega_{\rm cr}.  
\end{equation}
The differential rotation between the pinned superfluid in the nonlinear creep regime and the crustal normal matter results in energy dissipation at a rate
\begin{equation}
    \dot{E}_{\rm diss}\cong I_{\rm s}\omega_{\rm cr}|\dot\Omega|.
\label{nldis}
\end{equation}
Equation (\ref{nldis}) leads to a effective surface temperature
\begin{equation}
    T_{\rm s}=\left(\frac{\dot{E}_{\rm dis}}{4\pi R^{2}\sigma}\right)^{1/4},
\label{tsurf}
\end{equation}
where $R$ is the neutron star radius and $\sigma$ is the Stefan-Boltzmann constant. Equation (\ref{tsurf}) predicts that an old pulsar which radiated away its original heat content would shine in ultraviolet region of the thermal spectrum. The observations of a handful of pulsars confirm this prediction \cite{gonzalez10,becker06,zharikov08,durant12,rangelov17,guillot19,abramkin21,abramkin22}. Note also that other internal heating mechanisms, in particular rotochemical heating add constructively for high surface temperature detection in old pulsars \cite{gonzalez10,horvath22}.  
In nonlinear creep regions the steady state lag is very close to the critical threshold for vortex unpinning as pulsar ages and temperature drops accordingly. Small statistical fluctuations due to local temperature increase or shift due to a quake in the crust can easily raise the lag above the critical level so that the excess superfluid angular momentum is tapped by the collective unpinning and radially outward motion of vortex lines contained within. Kelvon waves induced on unpinned superfluid vortices and their coupling with the lattice phonons entails angular momentum exchange with ambient normal matter and thereby gives rise to a rapid speed up of the observed neutron star crust. The resulting angular momentum transfer process and therefore glitch spin-up time lasts for when these free vortex lines repin to new pinning sites. Therefore, these nonlinear creep regions are agent for pulsar glitches. At the time of a glitch superfluid rotation rate decreases by $\delta\Omega_{\rm s}$, i.e. $\Omega_{\rm s}\rightarrow\Omega_{\rm s}-\delta\Omega_{\rm s}$ and crustal rotation rate acquires an increment of amount $\Delta\Omega_{\rm c}$, i.e. $\Omega_{\rm c}\rightarrow\Omega_{\rm c}+\Delta\Omega_{\rm c}$. The glitch magnitude $\Delta\Omega_{\rm c}$ is then found from the angular momentum conservation at the time of the glitch event:
\begin{equation}
    I_{\rm c}\Delta\Omega_{\rm c}=I_{\rm s}\delta\Omega_{\rm s}.
\end{equation}
Thus, glitch magnitude is determined by two factors: the number of vortices unpinned at the time of glitch, $\delta N_{\rm V}=2\pi r^{2}\delta\Omega_{\rm s}/\kappa$ and the radial distance traversed by vortices before repinning. $I_{\rm s}$ consists of two parts: The nonlinear creep regions with moment of inertia $I_{\rm A}$ wherein vortices unpinned and regions with moment of inertia $I_{\rm B}$ through which vortices do not creep at all but contribute angular momentum balance via traverse of lines within them at the time of a glitch. Due to the movement of vortices across a glitch rotational energy of the pulsar changes by,
\begin{equation}
    \Delta E_{\rm rot}=\Delta J\omega_{\infty}=I_{\rm s}\omega_{\infty}\Delta\Omega_{\rm c},
\end{equation}
which is dissipated in the form of heat. Here $\Delta J$ is the angular momentum transfer across the glitch, $\omega_{\infty}$ is the pre-glitch lag between the pinned superfluid and the crust, $I_{\rm s}$ is the moment of inertia of the pinned superfluid and $\Delta\Omega_{\rm c}$ is the observed spin-up in the crustal rotation rate. The thermal luminosity expected from afterglow of glitches is given by \cite{vanriper91,cheng98,cheng01,cheng04}
\begin{equation}
    \Delta L=\frac{\Delta E_{\rm rot}}{\tau_{\rm th}},
\end{equation}
where $\tau_{\rm th}\approx10^{6}$ s is the thermal conduction time in the crust.

For constant external braking torque around the time of a glitch the post-glitch spin-down rate becomes
\begin{equation}
    \dot\Omega_{\rm c}(t)=\dot\Omega_{\infty}-\frac{I_{\rm cs}}{I_{\rm c}}\dot\Omega_{\infty}\left[1-\frac{1}{1+\left[\exp\left(\frac{t_{0}}{\tau_{\rm nl}}\right)-1\right]\exp\left(-\frac{t}{\tau_{\rm nl}}\right)}\right],
\label{creepsingle}    
\end{equation}
where we defined nonlinear relaxation time 
\begin{equation}
    \tau_{\rm nl}\equiv\frac{\varpi}{|\dot\Omega_{\infty}|},
\label{taunl}
\end{equation}
and offset time
\begin{equation}
    t_{0}\equiv\frac{\delta\omega}{|\dot\Omega_{\infty}|}.
\label{offtime}
\end{equation}

Eq.(\ref{creepsingle}) can be integrated explicitly by summing contributions of all non-linear creep regions into account with the assumption of linearly decreasing superfluid angular velocity during a glitch through a superfluid layer with radial extent $\delta r_{0}$, i.e. $\delta\Omega_{\rm s}(r, 0)=(1-r/\delta r_{0})\delta\Omega_{0}$ \citep{alpar96}

\begin{equation}
\Delta\dot\Omega_{\rm c}(t)=\frac{I_{\rm A}}{I_{\rm c}}\dot\nu_{0}\left(1-\frac{1-(\tau_{\rm nl}/t_{0})\ln\left[1+(e^{t_{0}/\tau_{\rm nl}}-1)e^{-t/\tau_{\rm nl}}\right]}{1-e^{-t/\tau_{\rm nl}}}\right).
\label{creepfull}
\end{equation}
Here total moment of inertia of the non-linear creep regions affected from vortex unpinning is indicated by $I_{\rm A}$. For times $\tau_{\rm nl}\lesssim t \lesssim t_{0}$ Eq.(\ref{creepfull}) reduces to
\begin{equation}
\frac{\Delta \dot \Omega_{\rm c}(t)}{\dot \Omega_{\rm c}}=\frac{I_{\rm A}}{I_{\rm c}}\left(1-\frac{t}{t_{0}}\right).
\label{glitchdotnu}
\end{equation}

After initial exponential recoveries are over the vortex creep model parameters are related to the post-glitch pulsar observables with three simple equations \cite{alpar06}:

\begin{equation}
    \frac{\Delta\Omega_{\rm c}}{\Omega_{\rm c}}=\left(\frac{I_{\rm A}}{2I}+\frac{I_{\rm B}}{I}\right)\frac{\delta\Omega_{\rm s}}{\Omega_{\rm c}},
\label{rotvc}    
\end{equation}
\begin{equation}
   \frac{\Delta\dot\Omega_{\rm c}}{\dot\Omega_{\rm c}}=\frac{I_{\rm A}}{I},
\label{drotvc}
\end{equation}
\begin{equation}
    \ddot\Omega_{\rm c}=\frac{I_{\rm A}}{I}\frac{\dot\Omega_{\rm c}^{2}}{\delta\Omega_{\rm s}}.
\label{ddrotvc}    
\end{equation}
The pre-factor $1/2$ in equation (\ref{rotvc}) accounts for the assumption of linear decrease of superfluid angular velocity through nonlinear creep regions with moment of inertia $I_{\rm A}$. Equations (\ref{rotvc}), (\ref{drotvc}) and (\ref{ddrotvc}) predicts time to the next glitch from the magnitudes of the previous one as
\begin{equation}
t_{0}=\frac{\delta\Omega_{\rm s}}{|\dot\Omega|}=2\times10^{-3}\left[\left(\Delta\Omega_{\rm c}/\Omega_{\rm c}\right)_{-6}/\left(\beta+1/2\right)\left(\Delta\dot\Omega_{\rm c}/\dot\Omega_{\rm c}\right)_{-3}\right]\tau_{\rm sd},
\label{toffset}
\end{equation}
where $\beta\equiv I_{\rm B}/I_{\rm A}$. The same set of equations gives an estimate for the interglitch braking index
\begin{equation}
    n_{\rm ig}=\left(\beta+1/2\right)\left(\Delta\dot\Omega_{\rm c}/\dot\Omega_{\rm c}\right)_{-3}^{2}/\left(\Delta\Omega_{\rm c}/\Omega_{\rm c}\right)_{-6}.
\label{nig}
\end{equation}
Equation (\ref{nig}) estimates as large as $n_{\rm ig}\sim100$ which compares well with the observed values \cite{ymh13,Dang2020,ljd21}.

Vortex creep model has been applied to post-glitch timing data of a handful of pulsars \cite{alpar84c,alpar93,chau93,akbal17,erbil20,alpar85b,alpar96,erbil19,alpar88b,akbal15,egy22}. From the vortex creep model fits to the post-glitch timing observational data many invaluable information probing into neutron star internal structure and dynamics can be obtained. Among them, total moments of the inertia of the superfluid regions participated to glitches provides the fractional moment of inertia of the crustal superfluid and in turn places constraint on the equation of state of neutron star matter. The superfluid recoupling timescale given by equation (\ref{taunl}) with an information on the temperature for a given pulsar helps to restrict related microphysics at subnuclear densities including superfluid pairing gap. The theoretical estimate of equation (\ref{offtime}) for the time to the next glitch agrees well with the observed interglitch time-scales especially for middle aged pulsars exhibiting Vela-like glitches \cite{egy22}. Therefore such estimates for other pulsars can be used to plan targeted observations of next glitches for the corresponding sources.       

Anti-glitches with reverse sign in the glitch magnitude can also be explained in terms of vortex avalanches under radially inward bias \cite{pines80,howitt22}. Crustquakes as an external agent may drive some of vortices radially inward at the time of glitch and acts as the required inward bias \cite{akbal15}. According to the vortex creep model delayed spin-ups observed in the young Crab pulsar results from the response of the internal superfluid torque on the neutron star crust when vortex lines transported inward after a crustquake \cite{erbil19}. Slow glitches, on the other hand, may be different manifestation of new vortex current distribution in old pulsars after moving inward due to crust breaking quake.    

Spin-down rate oscillations observed in the long term rotational evolution of some pulsars \cite{lyne10,kerr16,aditya19,shaw22} as well as post glitch relaxation of the Vela pulsar \cite{mcculloch90} and PSR B2234+61 \cite{yuan10,espinoza17} may result from the response of vortex line oscillations under the competition between the pinning potential and tension \cite{erbil22}.

If pulsar glitches are originated from crustal superfluid alone like in the vortex creep model, then the permanent post-glitch increase in the spin-down rate after exponential recoveries are over can be used to place limits on the crustal thickness and equation of state of neutron star matter and in turn pulsar mass \cite{datta93,link99,ho15,pizzochero17,montoli20,li21}. 

The tightest limit on the glitch rise time to date is obtained for the observation of 2016 glitch of the Vela pulsar and implies a spin-up of the neutron star less than 12 s \cite{ashton19}. Such short spin-up time-scales result from involvement of vortex lines via the coupling of kelvon modes induced on vortices with lattice phonons when they freed \cite{epstein92,jones92,graber18}. The relativistic extension of the mutual friction mechanism between the superfluid and normal matter \cite{sourie17,gavassino20}, and interpretation of the overshooting of the rotation rate immediately following the glitch as time variable coupling of neutron star core to the crust \cite{pizzochero20,erbil20,montoli20b} provide insights for neutron star internal structure and dynamics.

Vortex pinning to lattice nuclei is at the heart of many glitch models including the vortex creep model invoking pinned crustal superfluid as the angular momentum reservoir. A vortex will pin on to a nucleus if the interaction between them is attractive while hold on interstices of the lattice if the vortex-nucleus interaction is repulsive \cite{alpar77,link91}. As the neutron star crust spins-down under the combined action of the magnetic dipole radiation and pulsar wind emission, the vortex pinning stores angular momentum in some part of the interior superfluid component by fixing angular velocity of the pinned superfluid \cite{anderson75}. The amount of angular momentum maintained in the pinned superfluid which drives a glitch is determined by the strength at which vortex lines pin to lattice nuclei. It is shown that pinning at various strengths may occur depending on the interaction potential between vortex and nucleus, the lattice orientation with respect to the vortex line and the vortex tension \cite{jones97,seveso16,wlazlowski16,link22}. In particular an amorphous crustal structure or a lattice with large impurities would result in strong pinning wherein the unpinning process may start \cite{jones99,jones01}. A vortex is forced to move through lattice which will cause it to fall into interaction potential well provided by neighbouring nuclei, exiting waves on the vortex. As these waves damp as a result of the energy imparted to lattice phonons, the final shape of a vortex would be determined by the competition between gain due to being located at the bottom of the potential well and cost due to lengthening after bending \cite{bildsten89,epstein92,jones92,hirasawa01}. There has been substantial controversy in the literature concerning the strength of vortex-nucleus pinning interaction as well as the interaction sign. Quantum calculations of Avogadro et al. \cite{avogadro07,avogadro08} employ a mean field Hartree-Fock-Bogoliubov formalism approach to vortex-nucleus interaction and results in $\sim3$ MeV for the pinning energy per intersection. According to results of their calculations the interaction is in repulsive nature throughout sizeable portion of the inner crust. Pizzochero and his collaborators \cite{pizzochero97,donati03,donati04,donati06} have used local density approximation in their semi-classical calculations and obtained a repulsive interaction of magnitude $\sim1-2$ MeV below densities $\sim2\times10^{13}$~\,gr\,cm$^{-3}$ and an attractive interaction of strength $\sim5$ MeV at higher densities. An exact density functional theory approach to the same problem has yielded a repulsive interaction with $\sim4$ MeV per nucleus for the density range $(3-7)\times10^{13}$~\,gr\,cm$^{-3}$ \cite{wlazlowski16}. Jones \cite{jones97} argues that for an infinite, single component body centered cubic (bcc) lattice vortex line orientation with respect to different lattice planes would cancel out the pinning force on a vortex largely and concludes that vortex lines are in the co-rotation with the crustal lattice without pinning. However, the neutron star crust likely involves not such symmetric but multicomponent bcc lattice which renders pinning to be effective for specific lattice orientations \cite{grill12}. Also neutron star crustal lattice is not expected to be an ideal bcc as thermal fluctuations in the era of crustal solidification may lead to formation of monovacancies and lattice defects \cite{jones99,jones01}. It is speculated that above the neutron drip point the attractive force between the interstitial unbound neutrons and the lattice makes bcc lattice unstable \cite{kobyakov14}. However, when the Coulomb interaction between two nuclei is ignored the total attractive interaction is entirely due to exchange of phonons which results in cancellation of destabilisation effects \cite{kobyakov16}. Recently three dimensional simulations of vortex line-nuclei interaction have been performed to understand the vortex line-nucleus pinning problem and its implications for pulsar glitches \cite{link09,seveso16,antonelli20}. As the density increases lattice nuclei get closer so that they almost touch each other. The equilibrium shape of nuclei deviates considerably from being spherical as a result of destructive Coulomb and nuclear forces between them, and rod-like, plate-like nuclei appear. Such a sequence of nuclear shapes is collectively dubbed as pasta phase \cite{chamel08}. On the other hand, calculations of vortex pinning to non-spherical nuclei in the pasta phases of deeper layers of the neutron star crust has only premature nature \cite{lazzari95a,lazzari95b,deblasio96}.   

In the neutron star crust dissipationless coupling of the dripped superfluid neutrons with lattice nuclei restricts mobility of unbound neutrons and gives rise to a significant decrement for the effectiveness of superfluid neutrons to impart their angular momentum to the crust in glitches \cite{andersson12,chamel13,li15,delsate16}. Band theory calculations of \cite{chamel17} show that this entrainment effect may lead to large effective to bare mass ratio for neutrons with average enhancement factor $<m_{\rm n}^{*}/m_{\rm n}>=5$ throughout the neutron star crust. Therefore, the inferred moments of inertia found from post-glitch spin-down fits should be multiplied by the same enhancement factor. However, if the lattice in the neutron star crust is disordered, then the effects of the entrainment become less important \cite{sauls20}. Uncertainties regarding pairing potential may lead to smaller enhancement factors \cite{kobyakov16,urban16,watanabe17,durel18,watanabe22}. The pasta phases in the deepest regions of the inner crust with the series of different geometries of the atomic nuclei brings about reduction in the strength of entrainment compared to the spherical nucleus case \cite{sekizawa22}. Also some parts of the neutron star outer core may be involved in the glitches \cite{erbil14,montoli20}, which provided new way out for crustal superfluid based models.

So far we have investigated the effects of vortex pinning and creep for laminar superfluid flow. In the literature the effects of superfluid turbulence on pulsar glitches have been studied in some extent \cite{greenstein70,peralta05,peralta06,peralta07,andersson07,mongiovi17,haskell20}. Instabilities associated with the superfluid hydrodynamical flow were proposed to play an important role in initiating collective vortex unpinning cascade \cite{andersson03,mastrano05,sidery08,glampedakis09,andersson13,khomenko19}. It is shown that the creep and pinning of vortices across flux tubes in the outer core stabilises the superfluid flow and inhibits large scale turbulence \cite{vaneysden18}. 

\subsection{Crustquake Model}
Crustquakes are failures of the solid neutron star crust, resulting from spin-down assisted continual reduction in the centrifugal force which deforms and stresses the crust until it breaks. Crustquakes both spins up the star and dissipates mechanical energy inside it, thereby producing heat.

The outermost layer of a neutron star solidifies in an early epoch of the star's life when it was spinning comparatively faster and thus had a relatively oblate shape. As a neutron star spins down its fluid interior adjusts its shape to instantaneous new spin rate while its rigid crust resits such changes. Then, stresses build up in the crust which will be released in discrete quake events when the yield point is exceeded \cite{ruderman69,baym71}. The shape change after the crack reduces the moment of inertia of the crust with reference oblateness accompanying a jump. The conservation of angular momentum and torque equilibrium acting on the neutron star at the time of the glitch implies
\begin{equation}
    \frac{\Delta\Omega}{\Omega}=\frac{\Delta\dot\Omega}{\dot\Omega}=-\frac{\Delta I}{I}\approx2\frac{\delta R}{R},
\end{equation}
where $\delta R/R$ denotes the shrinkage in the solid component of the neutron star and is approximately given by \cite{ruderman69} 
\begin{equation}
\frac{\delta R}{R}\approx\frac{95}{7}\frac{\mu R}{GM\rho}\Delta\phi.
\end{equation}
Here, $R$ is the radius and $M$ is the mass of the neutron star, $\rho$ is the average stellar density, $\Delta\phi$ is the strain relieved in the quake and $\mu\approx(\Delta/R)(Ze)^{2}n^{4/3}$ is the shear modulus with $\Delta/R\sim0.1$ being the crustal thickness to neutron star radius ratio, $Ze$ being the nuclear charge and $n$ being the number density of nuclei. Recent theoretical and computational calculations show that the more accurate shear modulus value for rigid neutron star crust does not differ significantly from the above quoted order of magnitude estimate \cite{ogata90,strohmayer91,baiko11,chugunov22}. In Ref.\cite{alpar85} it has been argued that since the only dimensionless parameter related to the problem is the fine structure constant $\alpha=e^{2}/\hbar c$, the critical strain angle for crust cracking should be of the order $\theta_{\rm cr}=Ze\alpha$. The recent molecular dynamical simulations, which take realistic shear motion of the neutron star crust into account, yield $\theta_{\rm cr}\simeq0.04-0.1$ \cite{horowitz09,hoffman12,baiko18}, and is similar in order of magnitude to the early theoretical estimate.
Starquake model has a definite prediction for the repetition of glitches since the time between successive quakes is that required to build up the right amount of stress from its current value to the critical strain.

Baym and Pines \cite{baym71} established the formulation of starquake model for the solid neutron stars. In their model neutron star involves quadrupolar stellar distortion giving rise to mechanical stress. It is determined by a single time-dependent parameter, the crustal oblateness which can be only partially relieved during a quake. For quadrupolar deformation the rotational energy of a neutron star can be written as
\begin{equation}
E=\frac{L^{2}}{2I_{0}}+A\epsilon^{2}+B(\epsilon-\epsilon_{0})^{2},
\label{eqquake}
\end{equation}
where the oblateness $\epsilon$ is defined by the relation $I=I_{0}(1+\epsilon)$ with $I_{0}$ being the moment of inertia of the nonrotating spherical star and $\epsilon_{0}$ is the reference oblateness. In equation (\ref{eqquake}) $A$ and $B$ quantifies the gravitational potential energy and the elastic energy stored in the neutron star crust, respectively. Minimizing equation (\ref{eqquake}) with keeping $L$ and $\epsilon_{0}$ fixed gives
\begin{equation}
\epsilon=\frac{I\Omega^{2}}{4(A+B)}+\frac{B}{A+B}\epsilon_{0}.
\end{equation}
A quake on the crust occurs whenever the mean stress in the crust
\begin{equation}
\sigma=\frac{\partial E}{\partial\epsilon}\frac{1}{V}=\mu(\epsilon_{0}-\epsilon),
\end{equation}
with $V$ being the crustal volume exceeds the critical value $\mu\theta_{\rm cr}$. In a quake both $\epsilon$ and $\epsilon_{0}$ reduces according to
\begin{equation}
    \Delta\epsilon=\frac{B}{A+B}\Delta\epsilon_{0},
\end{equation}
where $\Delta\epsilon=\Delta I/I=-\Delta\Omega/\Omega$ is the observed glitch magnitude. After the quake stress will accumulate again once more and the next quake will take place after a time
\begin{equation}
    t_{\rm q}\approx\frac{\Delta\epsilon_{0}-\Delta\epsilon}{\dot\epsilon}\approx\frac{A}{B}\frac{\Delta\epsilon}{2\epsilon}\frac{\Omega}{\dot\Omega},
\end{equation}
with the time evolution of the oblateness is given by
\begin{equation}
    \dot\epsilon\approx\frac{\partial I}{\partial\epsilon}\frac{\Omega\dot\Omega}{4A}.
\end{equation}
There are also other means for sinks of stresses in the neutron star crust, most prominent being the plastic flow \cite{smoluchowski70,cheng92,gourgouliatos21}. If the response of the neutron star crust to the applied stress is not in the form of plastic flow the time to the next glitch is proportional to the amount of the stress relieved in the previous quake and is predictable for a given neutron star model \cite{cheng92}. The time interval between successive quakes is proportional to $\Delta\epsilon_{0}$ and set by critical strain angle at which solid crust exceeds the yield point. 

The small Crab-like glitches can be accounted for stiff neutron star equation of state while larger Vela-like glitches cannot be explained within the framework of the starquake model as $(A/B)(\Delta\epsilon/\epsilon)$ becomes larger than the typical interval between the glitches \cite{pines72,pines74}.

Since the rigidity parameter $b=B/A$ is so small, change of the ellipticity associated with a quake is tiny \cite{cutler03,zdunik08}. The actual value of the rigidity parameter $b$ is rather uncertain mainly reflected in computation of $B$. Franco et al. \cite{franco00} considered a uniform density solid crust having constant shear modulus afloat on a fluid core which led them to obtain 5 times smaller $b$ compared to Baym \& Pines model \cite{baym71}. Cutler et al. \cite{cutler03} obtained even smaller (50 times) rigidity parameter by taking compressible neutron star matter and equation of state dependent shear modulus into account but still working in Newtonian gravity. Since the glitch induced increase in the spin-down rate is proportional to $b$ the correct evaluation of structural parameters is vital in crustquake models which is far from being complete \cite{reisenegger21,gittins21,horowitz22}. 

In Baym \& Pines model the coefficients $A$ and $B$ were evaluated for a completely solid, incompressible constant density neutron star with uniform shear modulus in Newtonian gravity. However, the deformation of a realistic neutron star with different core and crustal densities is qualitatively dissimilar from the case of equal density incompressible star. In the following years the Baym \& Pines model extended to include the effects of general relativity \cite{carter75}, magnetic field \cite{franco00,kojima21}, compressible neutron star model with density dependent shear modulus \cite{ushomirsky00,cutler03}. 

Molecular dynamical simulations suggest large values of $\theta_{\rm cr}$ implying that the secular spin-down may only produce crustquakes just a few times in the whole life of a neutron star \cite{horowitz09,hoffman12,baiko18}.

Crustquakes can also play the role of trigger for vortex unpinning events by forcing vortices to leave their equilibrium configuration as their ends are anchored to the moving broken crustal platelet  \cite{akbal15,akbal18,erbil19}. Crustquakes turn some metastable mechanical energy into heat by dissipation. Such heating may mobilize a very large number of vortices in the inner crust, thus serving as trigger for collective vortex unpinning events \cite{link96}. The energy released in a crustquake induced glitch is always much larger than the superfluid originated glitch of the same size in which angular momentum transfer from a faster rotating superfluid in the inner crust to the normal matter dissipates rotational energy inside the star \cite{epstein88b}.

In the presence of a global stellar magnetic field the crust fractures asymmetrically and fault planes moves in a way such that the angle between magnetic dipole and rotational axes and torque acting on the neutron star both increases \cite{link92,link98,epstein00}. This is proposed as a viable explanation of persistent step increases observed after the Crab glitches \cite{wong01,lyne15,GZL2020,SKL2021}. 

Progressive failures in the solid crust suffice to account for at least the small glitches observed in pulsars of all ages. Giliberti et al. \cite{giliberti19,giliberti20} computed the deformation and the strain due to reduction in the centrifugal force in time by modelling the neutron star a solid crust floating above a constant density core. This is a very simplifying assumption since stable density stratification is expected to be present in realistic neutron stars \cite{braithwaite09}. Most recent works on stresses in the crust rely on quantifying the displacement field with centrifugal force changes as a part of the rotational evolution \cite{giliberti19,giliberti20,reisenegger21,gourgouliatos21}.

Elastically deformed crust under stresses motivates searches for detection gravitational wave emission from neutron stars \cite{owen13}. It was shown that stresses in the neutron star crust may sustain an ellipticity much larger than implied by observational upper limits for continuous gravitational wave \cite{reisenegger21}.

The crustquake model is also envisaged in terms of solid quark stars \cite{peng08,zhou14,lai18,wang21}. Quark stars are bound by the strong nuclear force rather than the gravitational force relevant for neutron stars. For pure quark matter the shear modulus is larger than neutron star matter so that one can obtain Vela like large glitches in typical repetition timescale of a few years without exceeding observational bounds put by surface temperature measurements. There are two types of glitches within the quake model scenario for quark stars: Type I glitches are bulk invariable events resulted from solely a change in stellar ellipticity and stellar volume does not alter so that the amount of energy release is insignificant. While Type II glitches are bulk variable events involving stellar shape change induced probably by accretion and accompanied by large amount of energy. Type I glitches are invoked to explain standard glitches whereas Type II glitches may account for magnetar glitches with radiative changes. In the quark star model slow glitches are explained as conversion of a collapsed superficial layer with a melted layer beneath the surface. Cooling of this viscous fluid layer depletes the accumulated stress and releases energy to effectuate slow rise of pulsar spin.

\subsection{Vortex Line-Flux Tube Interaction Model}

About a year or so after the birth of a neutron star, its core temperature is expected to fall below the transition value and the interior proton plasma becomes a superconductor \cite{ho15}. Since the flux expulsion timescale accompanying this transition is substantially long the magnetic field inside a neutron star core organises into tiny flux tubes of cross section $\Lambda\sim10^{-11}$\,cm each carrying a field of $B_{\rm c}~\sim10^{15}$\,G \cite{baym69a}. The number density of flux tubes
\begin{equation}
    n_{\Phi}=\frac{B}{\Phi_{0}}=5\times10^{18}\left(\frac{B}{10^{12}\,\mbox{G}}\right)\,\mbox{cm}^{-2},
\end{equation}
with $\Phi_{0}=hc/2e\cong2\times10^{-7}$\, G\ cm$^{2}$ being the unit of magnetic flux in each tube, far exceeds the number density of vortex lines given by
\begin{equation}
    n_{\rm V}=\frac{2\Omega_{\rm c}}{\kappa}=10^{5}\left(\frac{\Omega_{\rm c}}{100\,\mbox{rad s}^{-1}}\right)\,\mbox{cm}^{-2},
\end{equation}
in a typical neutron star. Therefore, intimately strong interactions between these two types of topological defects in superfluid-superconducting interior of a neutron star are expected on qualitative grounds. Another important feature of proton superconductivity inside neutron stars is that following the transition with the nucleation of magnetic field into flux tubes tensile force and in turn the stress associated with the existence of magnetic field increases enormously, $B^{2}/8\pi\rightarrow B B_{\rm c}/4\pi$ while the repulsion between the flux tubes diminishes greatly since in equilibrium configuration they are separated well apart \cite{easson77,ruderman91}. When a vortex line and flux tubes are within a distance of $\sim\Lambda$ these two types of structures would strongly interact with each other and the velocity difference between them is the main factor leading to distinct dynamical consequences for pulsar glitches. Radially expanding vortices in a spinning down neutron star either cut through or carry along with them flux tube array for each encounter.

Ruderman and his collaborators \cite{ruderman91,ruderman98} considered the effect of vortex line-flux tube pinning in the core of a neuron star on the crustal failure and resulting evolution for pulsar glitches. Several years ago Sauls \cite{sauls89} and Srinivasan et al. \cite{srinivasan90} speculated that on microscopic-scale density fluctuation related interactions between neutron superfluid vortex lines and proton superconductor magnetic flux tubes within a neutron star’s core could give rise to a strong coupling between the star’s magnetic field and its spin history and thereby may play an important role for neutron star magnetic field evolution and spin glitches, respectively. This seminal idea was later strengthened by including the effect of a neutron vortex line’s own strong magnetic field generated due to proton entrainment mass currents around them \cite{jones91a,mendell91}. Since the footpoints of flux tubes are anchored in the crust base, the associated stress on the rigid crust from moving flux tubes would grow as a result of the migration of the vortex line-flux tube conglomerate during secular spin-down. Then, either the crustal strain will ultimately exceed the yield point, or the core’s vortex lines will be forced to cut through
the core’s flux tubes. The stress associated with the flux tube squeezing the crust base leads to failure of the solid and break it when the following condition is met \cite{ruderman09}: 
\begin{equation}
    \frac{BB_{\rm c}}{8\pi}L^{2}\gtrsim\left(\mu\theta_{\rm cr}\Delta\right)L,
\end{equation}
where $\Delta$ is the crustal thickness and $L$ is the lengthscale on which crust cracks. The main uncertainties in the solid neutron star crustal properties are the maximum sustainable shear strain and the amount of sudden strain-relaxation if yield point is exceeded. An initially hot neutron star rapidly cools down below the temperature via neutrino emission and crust-solidification begins just 10 s after the birth. A key ingredient of the model is the relative velocity of a flux tube with respect to the that of a vortex line. The comparative velocity $v_{\rm c}$ at which expanding vortex line array can push the flux tubes arrangement through core electron-proton plasma with the maximum pace would be \cite{ruderman98} 
\begin{equation}
  v_{\rm c}=\beta\left(\frac{\Omega}{100\,\mbox{rad s}^{-1}}\right)\left(\frac{10^{12}\,\mbox{G}}{B}\right)10^{-6}\,\mbox{cm s}^{-1}. 
\label{vlft}
\end{equation}
Here the parameter $\beta$ depends heavily on the various properties of the neutron star core:
\begin{equation}
\beta=0.4\ln\left(\frac{\Lambda}{\xi}\right)\left(\frac{B_{\rm V}}{10^{15}\,\mbox{G}}\right)\left(\frac{B_{\rm c}}{10^{15}\,\mbox{G}}\right)\left(\frac{60\,\mbox{MeV}}{E_{\rm F_{\rm e}}}\right)\left(\frac{10^{36}\,\mbox{cm}^{-3}}{n_{\rm e}}\right),    
\end{equation}
where $\Lambda$ is the London penetration depth, i.e. the lengthscale over which magnetic flux decays for a flux tube, $\xi$ is the core radius of flux tube, $B_{\rm V}$ is the magnetic field attached to vortices, $B_{\rm c}$ is the field strength of each flux tube, $E_{\rm F_{\rm e}}$ is the Fermi energy of relativistic degenerate electrons and $n_{\rm e}$ is the number density of electrons. Note, however, that the form of the driving force used in derivation of Equation (\ref{vlft}) which leads to magnetic field diffusion in the neutron star core has been questioned in the literature \cite{ding93,jahanmiri00,jones06,bransgrove18}.  

The vortex line-flux tube pinning model predicts two distinct families of glitches for pulsars. The diamagnetic screening currents set after superconducting transition in the neutron star core allows flux tubes to move independently through electron-proton plasma which leads bunching for flux tubes when they pushed by vortex lines \cite{ruderman03}. This is the underlying reason why the flux tubes move with the vortex lines rather than remaining immobile inside neutron star core and being cut through or carry along by the radially migrating vortices during spin-down of the neutron star. In young neutron stars since the radially outward vortex velocity is high crust cracking should also give rise to a permanent offset in the perpendicular dipolar component of global magnetic field configuration with respect to the spin axis. This would lead to spin-down rate change part of which does not heal back. In younger and thus more spinning down neutron stars like the Crab pulsar the horizontal motion of broken platelet relieves strain of amount $\Delta\theta\sim s/R\ll\theta_{\rm max}\sim10^{-1}$ with $s$ being the horizontal displacement of broken platelet. This sudden reduction in the shear stress on the crust by the flux-tubes just beneath it has a magnitude $(B B_{\rm c}/8\pi)(s/R)$. Thus, the glitch magnitudes in spin and spin-down rates are related by \cite{ruderman05}
\begin{equation}
    \frac{\Delta\Omega}{\Omega}\sim\frac{B B_{\rm c}}{8\pi\rho R^{2}\Omega^{2}}(s/R)\sim10^{-4}\frac{\Delta\dot\Omega}{\dot\Omega}.
\end{equation}

As a pulsar ages (like the Vela pulsar) a different kind of glitch family emerges. If there were not very dense array flux tubes around vortex lines, the outward moving vortices would smoothly shorten as they approach to the neutron star equator and then disappear. However, the very existence of flux tubes prevents this to occur and vortices pile up inside an annulus neighbouring to stellar equator and thereby excess angular momentum is deposited there. Highly conductive crust resists the entry of flux tubes towards crust-core interface. When the Magnus force on these accumulated vortex lines overwhelms the repulsive force between closely packed flux tubes a sudden transfer of angular momentum from vortices to crust takes place making a spin up glitch. As a pulsar matures, the volume occupied by comoving vortex line-flux tube network increases and the glitch size attains much larger magnitudes given by \cite{ruderman05}
\begin{equation}
    \frac{\Delta\Omega}{\Omega}\sim\left(\frac{5V_{\rm A}}{2V_{\rm NS}}\right)\left(\frac{\tau_{\rm g}}{\tau_{\rm sd}}\right),
\end{equation}
where $V_{\rm A}$ is the volume of the annulus of the vortex line-flux tube interaction region and $V_{\rm NS}$ is the volume of the neutron star. The repetition time between successive Vela like giant glitches is given by \cite{ruderman05}
\begin{equation}
    \tau_{\rm g}\sim3\times10^{-4}\left(\frac{\Omega}{\dot\Omega}\right)|n-3|^{-1}.
    \label{tgrud}
\end{equation}

Equation (\ref{tgrud}) predicts $\tau_{\rm g}=0.7$ yr for $\tau_{\rm sd}=5$ kyr old "Big Glitcher" PSR J0537--6910 which has interglitch braking index $n_{\rm ig}\sim7$ and observed glitch interval $\tau_{\rm ig}=0.4$ yr \cite{middleditch06,ferdman18,antonopoulou18,ho20,abbott21,ho22}.

Flux tube-vortex line pinning model predicts that after each glitch the amount of the spin-down rate that does not heal is given by \cite{ruderman98}
\begin{equation}
    \frac{\Delta\dot\Omega_{\rm p}}{\dot\Omega}=\frac{(3-n)}{2}\frac{\tau_{\rm g}}{\tau_{\rm sd}},
\end{equation}
where $\tau_{\rm sd}=\Omega/(2|\dot\Omega|)$ is the pulsar spin-down (characteristic) time-scale. 

Delayed spin-up observed in several glitches of the Crab pulsar can be interpreted in terms of the vortex line-flux tube pinning model as the process of setting of vortex lines on new equilibrium position responding to unbalanced force on vortices due to retarding drag by flux tubes immediately after crust breaking.    

According to this model, the glitch variety observed in individual and different pulsars may be related to the complexities associated with breaking properties of the stratified neutron star crust \cite{ruderman09}. The various crustal layers with specified nuclei would have different lattice orientations with respect to the crystal symmetry axis. The vortex pinning strengths would be also altered by the crustal depth and the temperature evolution through the crust which further complicates the place where a lattice crack starts. All these effects add constructively to the diversity of observed glitches across pulsar population.

In all comparisons of observations with model-based predictions no apparent disagreement have been found. The application of the vortex line-flux tube pinning model to the post-glitch relaxation observations is discussed in Refs. \cite{jones02,wang00}.

Both the frictional drag of flux tubes within electron-proton plasma and the inelastic cut-through or pinning with vortex lines generate heat which can affect the thermal evolution of a neutron star \cite{sedrakian93,erbil17a}. The resulting heat production rate puts stringent limits on the vigour of the vortex line-flux tube pinning model \cite{ruderman98}.

Much of the important model ingredients like how vortex lines move together with flux tubes, under what
circumstances they cut through them, the associated drag forces acting on moving flux tubes and how the crust breaks under flux tube stress, remain to be investigated at more quantitative level.

\section{Radiative Changes Associated with Glitches}
Pulsar glitches are generally accepted to be caused by the interior dynamics of a neutron star \cite{baym69,anderson75,pines85}. If radiative changes (including long-lived flux enhancements, short bursts, and pulse profile changes) are associated with glitches, a link between the interior and magnetospheric state change would be naturally established. Based on available observational studies, concurrent variations in emission and the rotation of pulsars are a rather rare phenomena. Table \ref{emro} lists the detailed parameters of each observed correlation between glitch and emission change. It is seen that just a handful of normal pulsars display radiative changes that are concurrent with the glitch events. The emission change-rotation correlation is found to occur in both young and relatively old pulsars. Glitches in young mode-switching pulsars have typically large $\Delta\nu/\nu$ sizes, which follows the statistical trend \cite{basu22}. 

\begin{figure}
\centering
\subfigure{
\includegraphics[width=0.45\textwidth]{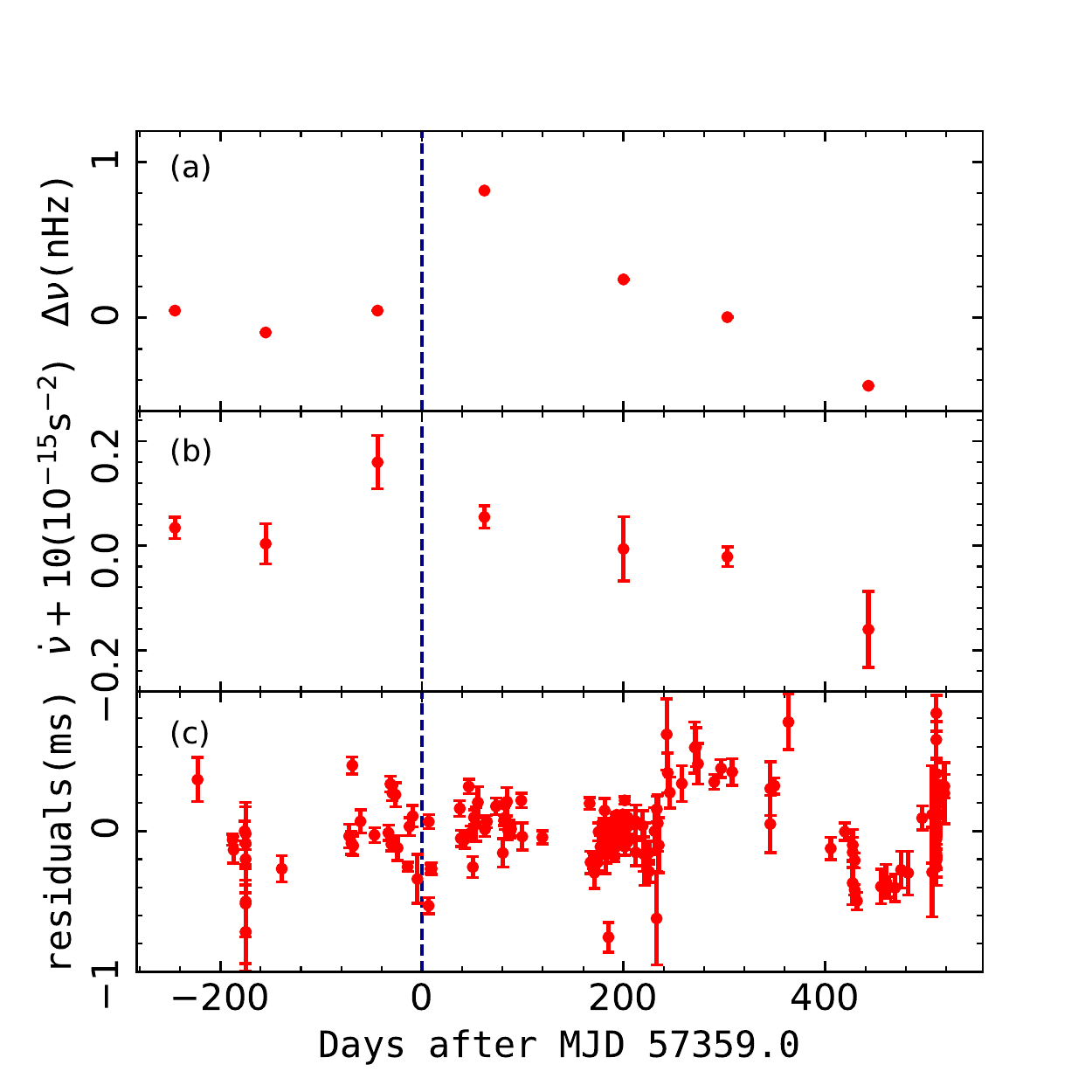}
}
\subfigure{
\includegraphics[width=0.45\textwidth]{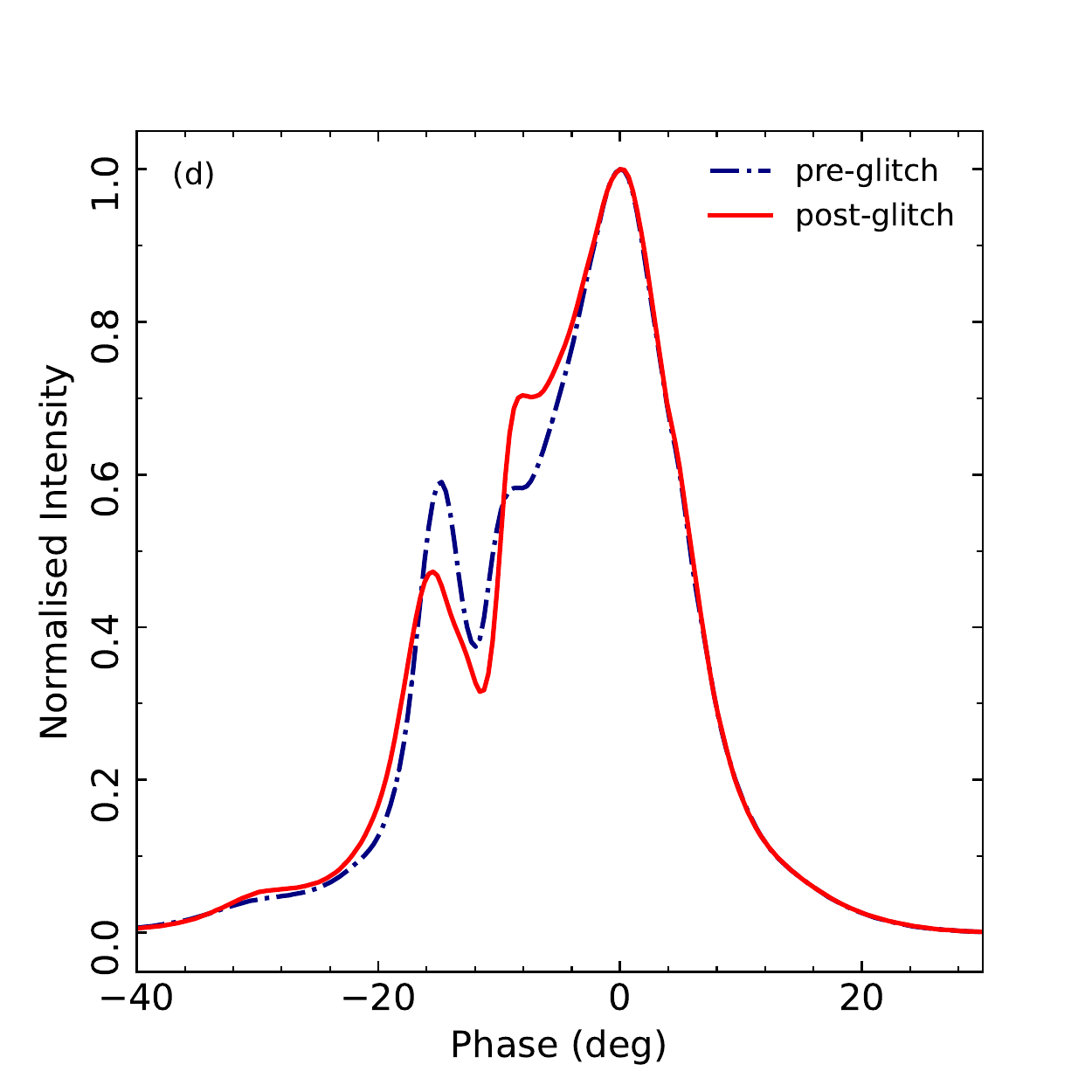}
}
\caption{Emission-rotation correlation in PSR J0738--4042 reported by Zhou et al. \cite{Zhou2022}: (a) variations of the frequency resuduals $\nu$; (b) the variation of $\dot{\nu}$; (c) the post-fit residuals after adding the glitch terms to the timing model. (d) the integrated normalized pulse profiles at pre- (blue line) and post-glitch (red line) modes. The vertical line indicates the glitch epoch at MJD $\sim57359(5)$.}
\label{0738}
\end{figure}

Recently, Zhou et al. \cite{Zhou2022} reported that a micro-glitch ($\Delta\nu/\nu \sim 0.36(4)\times10^{-9}$) occurred in middle-aged pulsar PSR J0738--4042 is coincided with variations in the power of the average pulse profile components. As demonstrated in Figure \ref{0738}, the average pulse profile after the glitch is slightly wider than the pre-glitch one, with a reduction in the leading component and an enhancement of the middle component. The post-glitch $\dot{\nu}$ shows a steady increase during this correlation, and similar post-glitch features have been seen in a very large glitch ($\Delta\nu/\nu \sim 33250\times10^{-9}$) in PSR J1718--3718 \cite{Manchester2011} and a delayed spin-up event in the Crab pulsar \cite{SLS2018}. The very young pulsar, PSR J1119--6127, is another intriguing example of mode-changing radio pulsars. A large glitch activity in PSR J1119--6127 with two exponential recoveries is linked to the appearance of additional pulse components that have an intermittent and RRAT-like behaviour \cite{Weltevrede2011}. Shortly afterward, Archibald et al. \cite{Archibald2016} observed another large glitch to be accompanied by X-ray bursts and X-ray pulsations in this pulsar. The outburst timing anomalies phenomena seen in PSR J1119--6127 are similar to those in magnetars, providing evidence for a connection between the high magnetic field radio pulsar and magnetar populations. The potential future for high-magnetic-field, rotation-powered pulsars is to turn into magnetars \cite{Espinoza2022}. It is important to make timing observations more frequently for high-B young glitching pulsars, such as PSRs J1718--3718 \cite{Zhu2011} and J1734--3333 \cite{Olausen2013}, as they might display magnetar-like mode-switching behaviours in coming years. 
The 2016 Vela glitch was accompanied by polarization changes and emergence of a null pulse \cite{palfreyman18}. It is proposed that changes in the state of magnetospheric emission reflects Alfven mode conversion close to the light cylinder triggered by energy release aftermath of a crustquake \cite{bransgrove20,yuan21}. 


\begin{table}[H]
\begin{center}
\begin{threeparttable}
\caption{Detailed parameters for pulsars exhibited correlation between glitches and emission changes.}
\begin{tabular}{lllcccrcc}
  
    \hline
    
\multicolumn{1}{c}{Pulsar Name}                  & \multicolumn{1}{c}{$P$}               & \multicolumn{1}{c}{$\tau_{c}$}                                  & \multicolumn{1}{c}{$B_{s}$}           & \multicolumn{1}{c}{$\dot{E}$} 			                         & \multicolumn{1}{c}{$\Delta\nu/\nu$}   & \multicolumn{1}{c}{$\Delta\dot{\nu}/\dot{\nu}$}                            & \multicolumn{1}{c}{Profile}           & \multicolumn{1}{c}{Ref.}                         \\

\multicolumn{1}{c}{(PSR)}                        & \multicolumn{1}{c}{(s)}               & \multicolumn{1}{c}{(kyr)}                                                  & \multicolumn{1}{c}{($10^{12} \rm\ G $)}& \multicolumn{1}{c}{($10^{32}\rm \ erg/s$)}                      & \multicolumn{1}{c}{($10^{-9}$)}       & \multicolumn{1}{c}{($10^{-3}$)}                                 & \multicolumn{1}{c}{}                  & \multicolumn{1}{c}{}                             \\
        
    \hline
    
    J0738--4042      & 0.3749                & 4320                   & 0.727                & 10
                    & 0.36(4)               & 3(1)                                      & $W_{50}$($\downarrow$)$^a$                       & \cite{Zhou2022} \\

    J0742--2822      & 0.1667                & 157                    & 1.69                 & 1400
                    & 102.73(11)            & 2.1(5)                                    & 
$^b$                                      & \cite{Keith2013, Dang2021}  \\

    J1119--6127      & 0.4079                & 1.61                   & 41                   & 23000
                    & 9400(300)             & 580(14)                                   &
$^c$                                          & \cite{Weltevrede2011}      \\

                    &                       &                        &                      & 
& 5740(80)         &  79(25)                          
&$X$($\uparrow$)              & \cite{Archibald2016}  \\

    B1822--09        & 0.7690                & 233                    & 6.42                 & 45
       & {{4.08(2)}} &  {{0.08(1)}}                         
&$^d$               & \cite{Liu2022}   \\

                     &                       &                        &                      & 
& {{7.2(1)}} &  {{1.65(7)}}                           
&$^d$               & \cite{Liu2022}   \\

    J2021+4026       & 0.2653                & 76.9                   & 3.85                 & 1200
                    & <100(--)          & 56(9)                  & $\gamma$($\downarrow$)
                                         & \cite{Allafort2013, Takata2020} \\

    B2021+51         & 0.5291                & 2740                   & 1.29                 & 8.2
                    & 0.373(5)              & --0.24(3)                                 & 
$W_{\rm 10}$($\downarrow$)$^a$                   & \cite{Liu2021} \\

    B2035+36         & 0.6187                & 2180                   & 1.69                 & 7.5
                    & 7.7(8)                 & 67(8)                                     & 
$W_{50}$($\downarrow$)$^a$                       & \cite{Kou2018}   \\


                    
    
     \hline

\end{tabular}
\begin{tablenotes}
\item[] $^a$: $W_{\rm 10}$ and $W_{\rm 50}$ are the full widths of the pulse profile at $10\%$ and $50\%$ of the peak pulse amplitude, respectively.\\
$^b$: represents that the appearance of additional pulse components is closely correlated with an unusual glitch.\\
$^c$: indicates the correlation between the ratio of the two-components in the profile and $\Delta\dot{\nu}$, which rapidly increases after the glitch.\\
$^d$: implies the variations of the integrated mean pulse profiles in both the radio-bright (B-mode) and the radio-quiet (Q-mode) modes.\\
\end{tablenotes}
\label{emro}
\end{threeparttable}
\end{center}
\end{table}

The mechanism behind these state-switching correlations is unclear. Despite this, the scenario that the glitch may change the magnetic field structure and hence the inclination angle $\alpha$ \cite{akbal15, Ng2016, Kou2018}, is gradually becoming more accepted. The outcome of this event is a change in the pulsar emission profiles and  spin-down torque. In the cases of PSRs B1822--09 and B2021+51, Liu et al. \cite{Liu2021, Liu2022} proposed that the flux tube, which moves independently in the emission zone, changes position during a glitch, and as a result the pulse profile varies. Zhou et al. \cite{Zhou2022} performed an in-depth analysis for the simultaneous variations in profile shape and spin-down rate $|\dot{\nu}|$ in PSR J0738--4042, based on a combined model of crustquake induced platelet movement and vortex creep response. They evaluated that the glitch-triggered emission variations may be a result of the crustquake-induced change in the inclination angle and the movement of vortex lines across the glitch \cite{Zhou2022}.


\section{Statistics of Glitches}
\begin{figure}[H]
\centering
\includegraphics[width=0.75\textwidth]{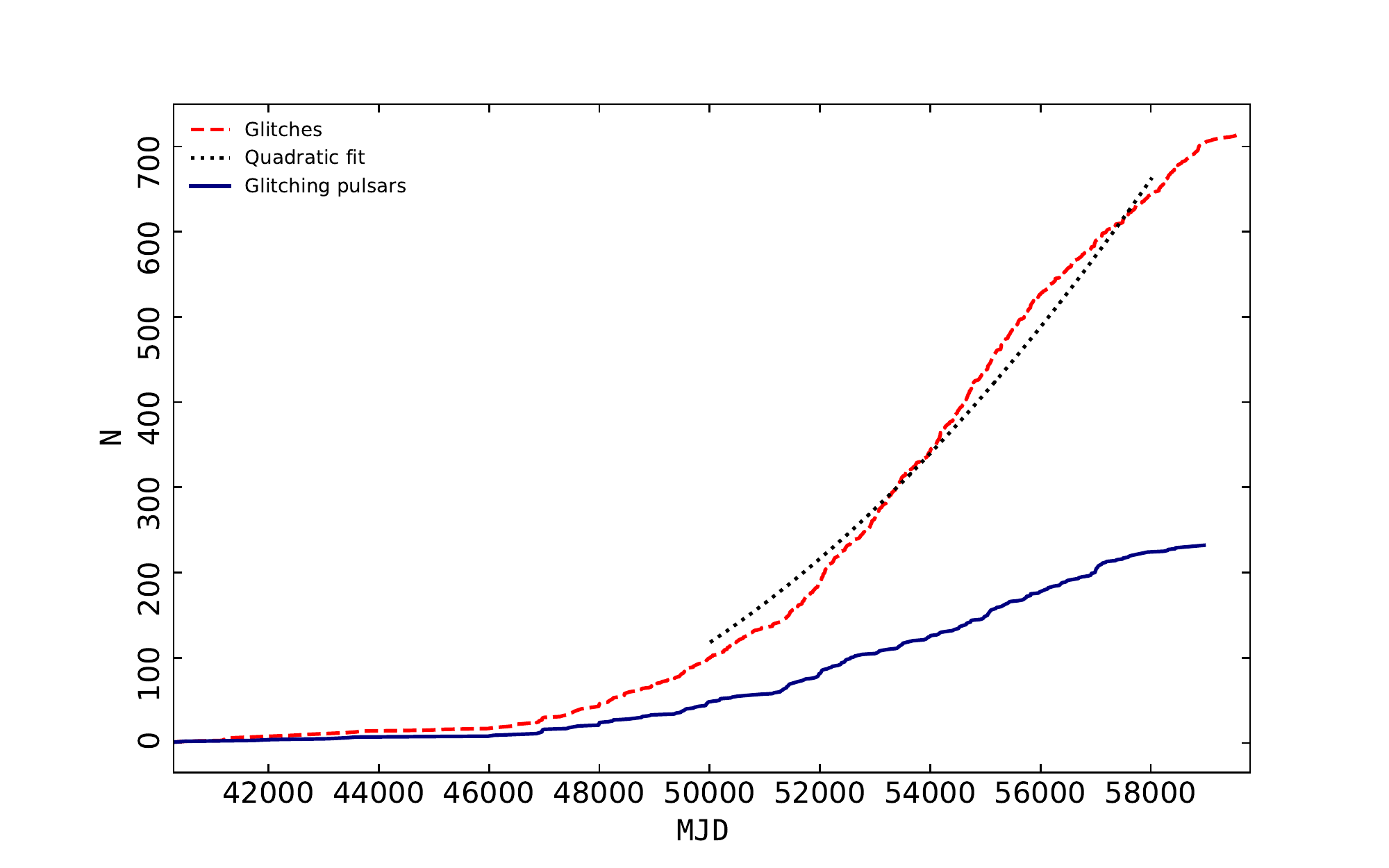}
\caption{Cumulative number of the detected glitching pulsars (blue) and glitches (red) over time. A quadratic fit (black) to the glitch discovery $N \propto t^2$ for $50000 < \rm MJD < 58000$ is carried out. The discovery epochs of glitching pulsars and glitches are collected from the combined glitch catalogues at ATNF \textsuperscript{\ref{foot1}} and Jodrell Bank (JBO) \textsuperscript{\ref{foot2}}.}
\label{Noofglitches}
\end{figure}
After they have been observed in sufficient numbers of pulsars since their first discovery in 1969, there have been numerous statistical studies on the magnitudes, repetition times and relaxation behaviours of the pulsar glitches \cite{alpar94,lyne00,wang10,espinoza11,ymh13,eya19,montoli21,basu22,egy22,eya22}. The total number of glitches occurred and glitching pulsars over time is shown in Figure \ref{Noofglitches}. With increased detection of new pulsars and follow up observations of more pulsars by many radio telescopes, a steady big increase in the number of pulsar glitches has been witnessed over the past 50 years. A curve fit is applied to find the cumulative number of the glitch discovery between MJDs 50000 and 58000 in Fig. \ref{Noofglitches}, obtaining a quadratic function $N \propto t^2$. Several statistical studies in particular datasets have been conducted to reveal the glitch populations and activities, even though the existing datasets suffer from incompleteness due to various reasons.

\subsection{Glitches Sizes Distributions $\&$ Waiting Times}
\begin{figure}[H]
\centering
\includegraphics[width=0.8\textwidth]{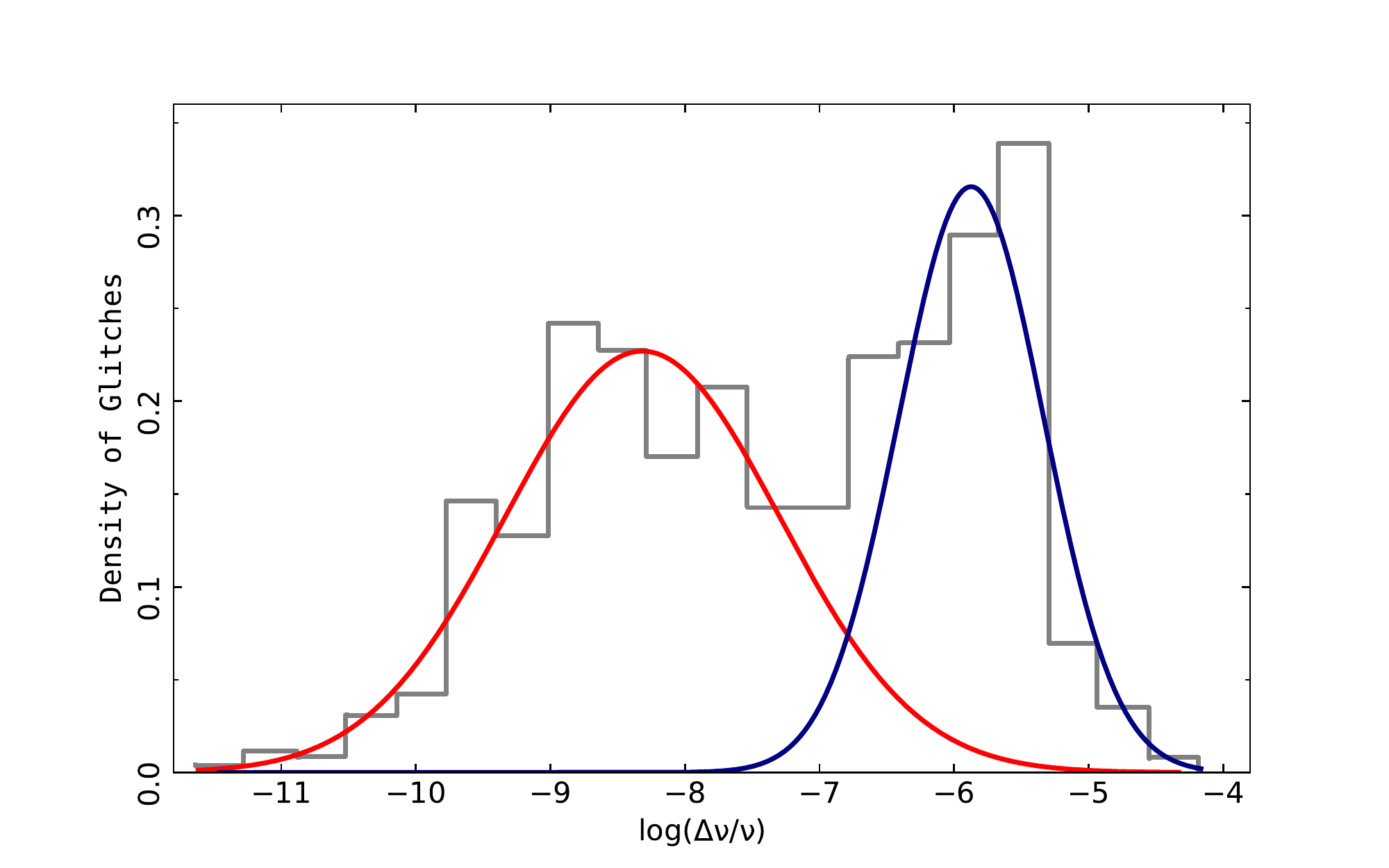}
\caption{Normalized histogram of the pulsar glitches amplitudes log($\Delta\nu/\nu$) \cite{Arumugam2022}. There are 390 glitches which belong to the first component and 310 to the second component. The narrower Gaussian component (navy) centred at $1.3\times10^{-6}$ represents the large glitches, whereas the wider Gaussian component (red) centred at $5.9 \times 10^{-9}$ represents the smaller glitches.}
\label{distribution_glitch}
\end{figure}

It was pointed out by Espinoza et al. \cite{espinoza11} that both the distributions of absolute and fractional frequency jumps, $\Delta\nu$ and $\Delta\nu/\nu$, of all observed glitches show the bimodal nature. This behaviour had also been indicated from recent studies using a larger sample \cite{ymh13, fuentes17, Ashton2017, basu22, ljd21}. Basu et al. \cite{basu22} used the best-fit two-component Gaussian Mixture Model to label the $\Delta\nu$ distribution using 543 glitches in 178 pulsars. The first, wide Gaussian component with small magnitudes centres at $\Delta\nu\sim0.032\ \mu \rm Hz$ and the second, narrow Gaussian component with larger magnitudes at $\Delta\nu\sim18\ \mu \rm Hz$. With a total sample of 700 glitches, Arumugam and Desai \cite{Arumugam2022} applied Extreme Deconvolution (XDGMM) method, which is a generalization of Gaussian Mixture Model, to classify the relative glitch amplitudes $\Delta\nu/\nu$ into two classes. Two pulsar glitch components of the bimodal distribution are ascertained, with the mean values  $5.9 \times 10^{-9}$ and $1.3\times10^{-6}$ (Figure \ref{distribution_glitch}), respectively. Moreover, there is a pronounced dip at $10^{-7}$ presented in the $\Delta\nu/\nu$ distribution, suggesting that the bimodality of the glitches sizes distribution may be produced by two mechanisms by which the glitch events occur \cite{ymh13, fuentes17}. As previously described in Section \ref{models}, largest glitches are caused by the sudden transfer of angular momentum from the faster rotating interior superfluid to the solid crust, whereas small glitches may be result from starquakes due to relaxation of the neutron-star crustal oblateness to the current equilibrium shape. Celora et al. \cite{celora20} proposed the non-linear mutual friction force as a cause of the exchange of angular momentum between the neutron superfluid and the observable normal component in glitches, and this effect is able to explain the observed bimodal distribution. An alternative explanation is that two different classes of pulsars may be the underlying reason of the bimodal behaviour \cite{Haskell2018}. As for spin-down rate changes $\Delta\dot{\nu}$, negative values are seen in the majority of glitches (that means after each glitch the spin-down rate becomes more negative since its absolute magnitude $|\dot\nu_{0}+\Delta\dot\nu|$ is greater), and there is usually a greater change in $\Delta\dot{\nu}$  for larger glitches \cite{basu22}.

For most pulsars glitches take place at irregular intervals with random distribution without a preferred distribution of magnitudes. The exceptions are the Vela pulsar and PSR J0537--6910 which have preferred large sizes and regular repetition times obeying Gaussian distribution \cite{fuentes19}. At the other extreme, the Crab pulsar glitches display temporal clustering at times \cite{lyne15,carlin19c}. The random distribution of glitch sizes and waiting times until the next event can be understood within the framework of the vortex creep model. As can be seen from equations (\ref{rotvc}) and (\ref{toffset}) both of the observed parameters depend on the two ingredients: number of vortices unpinned at the time of the glitch and the distance covered by them through inner crust which in turn determined by vortex scattering of pinning traps \cite{pines80,erbil20}. Shallow heating of crust via a quake may sustain different initial conditions for such scattering events. Even though inter-glitch time-scales and glitch sizes in a given pulsar differs significantly, the process may be regarded as scale invariant in terms of the sandpile effect. Radio pulsar glitches thus can be used to probe the far-from-equilibrium processes involving stress accumulation and relaxation in neutron star interiors \cite{carlin20,carlin21}. For productive pulsars exhibiting frequent large glitches the distribution of waiting times is found to be fitted by exponentials while their glitch size distribution obeys power law or log-normal functions \cite{melatos08,fulgenzi17,howitt18,melatos18,melatos19,fuentes19,carlin19a,carlin19b}.

\subsection{Glitch Activity}
The number of glitches per year $\dot{N}$ is a way to simply quantify how often glitches occur for known glitching pulsars \cite{espinoza11, ymh13, basu22}. According to Lyne et al. \cite{lyne00} and Espinoza et al. \cite{espinoza11}, the observed $\dot{N}$ is significantly correlated with characteristic age $\tau_c$ and spin-down rate $|\dot{\nu}|$. Basu et al. \cite{basu22} found that the maximum and minimum glitch rates for the 134 pulsars observed by the Jodrell Bank Observatory for a long time, were $\dot{N}_{\rm max}=1.07\ \rm yr^{-1}$ and $\dot{N}_{\rm min}=0.02\ \rm yr^{-1}$, respectively. The glitch activity trends in previous work were also confirmed by Basu et al. with a larger sample of glitching pulsars. $\dot{N}$ decreasing for pulsars with larger $\tau_c$ implies that young pulsars exhibit glitches more often than old pulsars. A higher $\dot{N}$ occurring alongside a greater $|\dot{\nu}|$ reinforces the notion that glitches are driven by the spin-down. However, this analysis does not take into account the size of the glitches, only the number; and the incomplete observations of glitches could skew these results. Thus, the integrated glitch activity is defined to objectively determine the cumulative effect of the glitches on the pulsar's rotation \cite{lyne00, fuentes17, basu22}:
\begin{equation}
\dot{\nu}_{\rm g}=\frac{\sum_{i}\Delta\nu_{i}}{T},
\label{glitchac}
\end{equation}
where $\Delta\nu_{i}$ is the spin frequency increment at the $i$-th glitch, and $T$ is the duration over which the pulsar has been searched for glitches. It is  intriguing how the integrated glitch activity correlates with key pulsar quantities, such as spin-down rate $|\dot{\nu}|$, spin-down age $\tau_{c}$, energy loss rate $\dot{E}$, and magnetic field $B$. To carry out a robust estimation for these correlations, the average glitch activity $\overline{\dot{\nu}}_{\rm g}$ for groups of pulsars  with similar properties is examined \cite{fuentes17,basu22}. The specific results are: the linear relation between $\overline{\dot{\nu}}_{\rm g}$ and $|\dot{\nu}|$ in the range $-14 < \log|\dot{\nu}| < -10.5$
 is  $0.018(3)$, implying that $|\dot{\nu}|$ undergoes a reversal of about 1.8 percent through glitches; $\overline{\dot{\nu}}_{\rm g}$ decreases with $\tau_{c}$; $\dot{E}$ are correlated positively with $\overline{\dot{\nu}}_{\rm g}$; pulsars with lower $B$ also have lower $\overline{\dot{\nu}}_{\rm g}$, and an increase in $\overline{\dot{\nu}}_{\rm g}$ is seen in neutron stars with the strongest $B$ (magnetars) \cite{lyne00,espinoza11,fuentes17, basu22}.

\begin{figure}[H]
\centering
\includegraphics[width=0.45\textwidth]{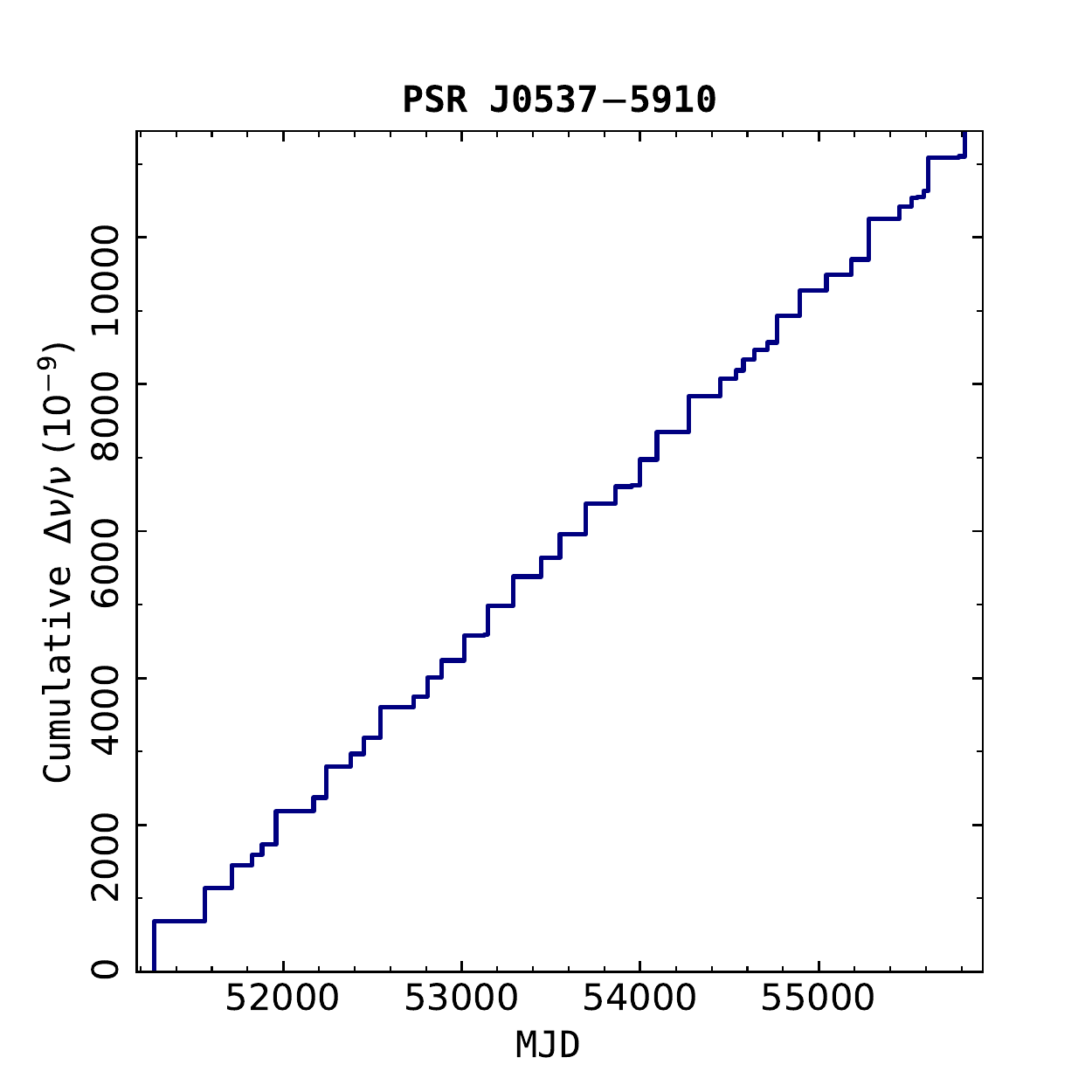}
\includegraphics[width=0.45\textwidth]{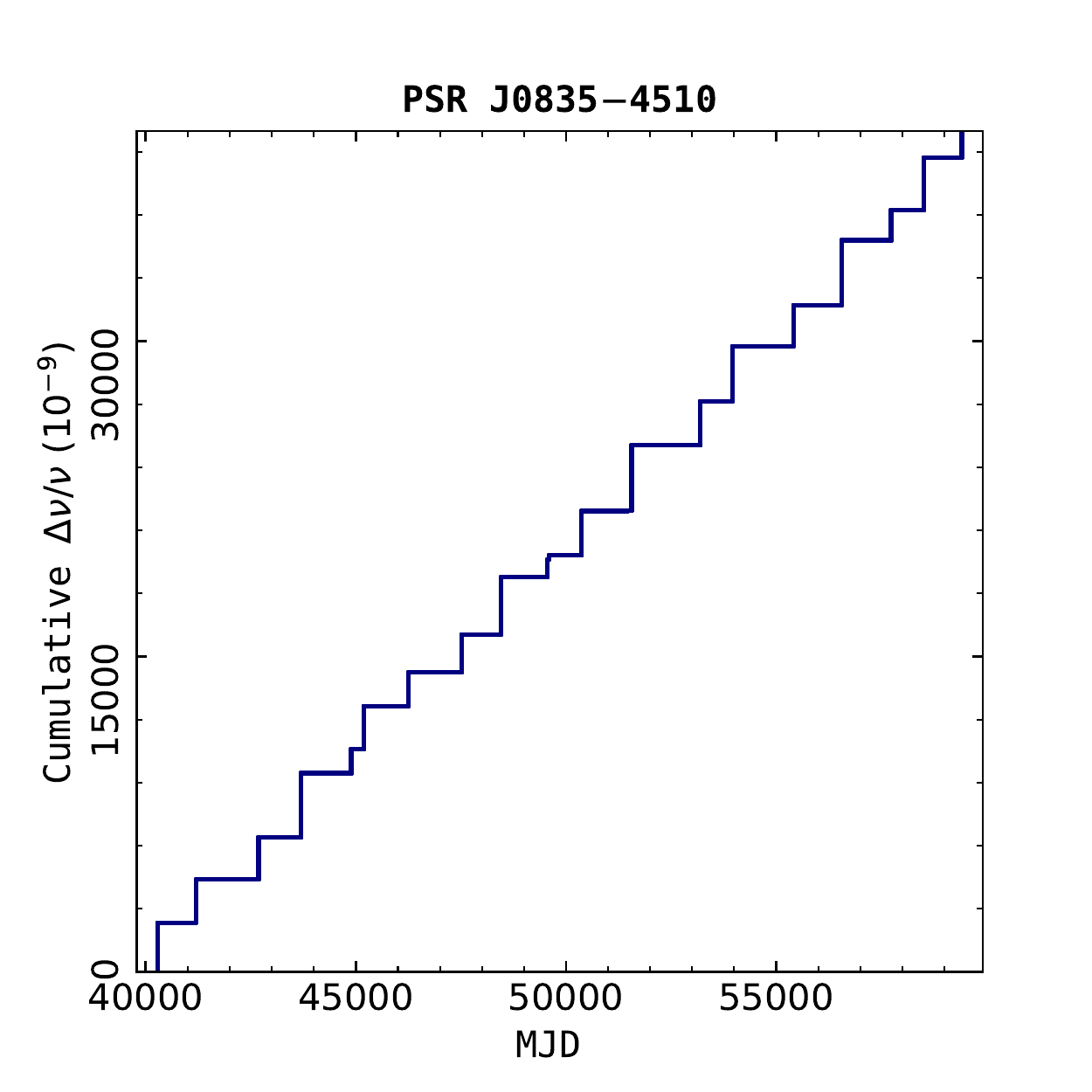}
\includegraphics[width=0.45\textwidth]{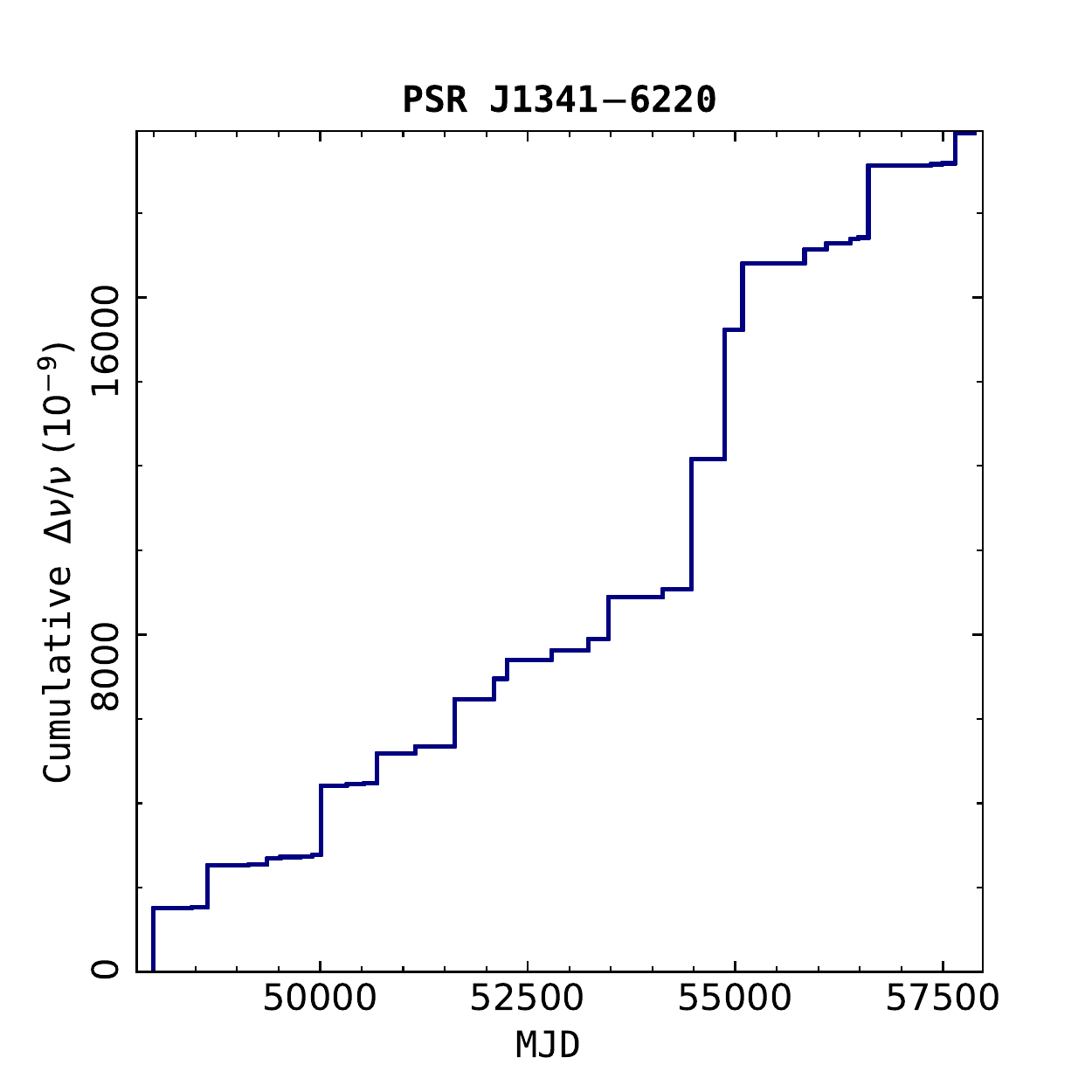}
\includegraphics[width=0.45\textwidth]{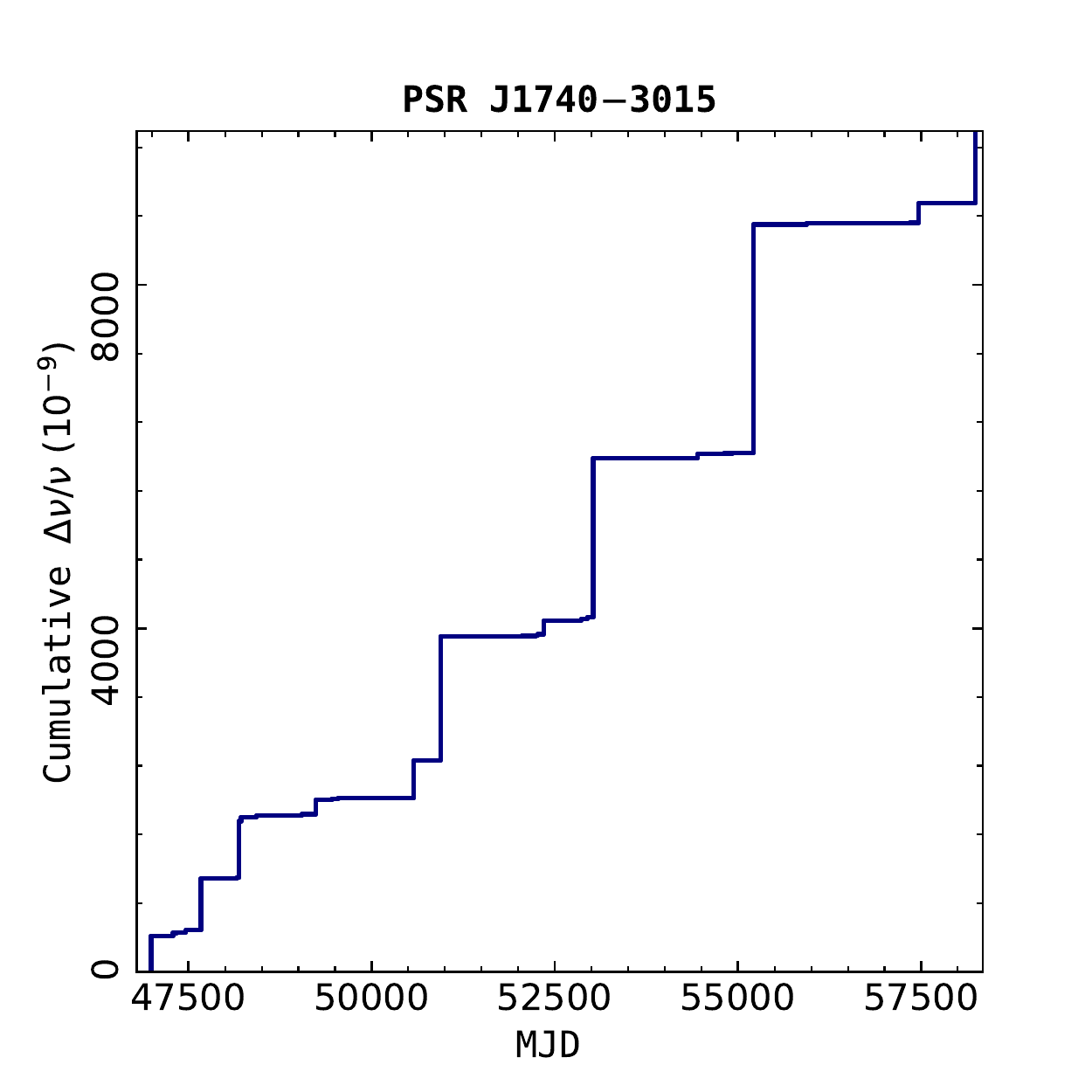}
\caption{The accumulated $\Delta\nu/\nu\ (10^{-9})$ for the most frequent glitching pulsars PSRs J0537$-$6910, J0835$-$4510 (Vela), J1341$-$6220 and J1740$-$3015.}
\label{cumulative}
\end{figure}

The accumulated effect of spin-up glitches on the rotational evolution for the long-term observations of PSR J0537--6910, the Vela pulsar (PSR J0835--4510), PSR J1341--6220 and PSR J1740-3015 are shown in Figure \ref{cumulative}. The step-like increases correspond to glitches and imply that spin-up contributions are not totally reversed by the ongoing secular slow-down due to electromagnetic dipolar radiation and pulsar wind. In terms of the vortex creep model \cite{alpar84a,cheng88,erbil20} this observation corresponds to the fact that at the time of a glitch vortex lines move through vortex traps inside which vortices do not creep at all. Since inside these traps continues angular momentum impart between glitches does not occur via creep process, the associated part of spin-up increase as a result of discrete angular momentum transfer by glitches persists for interglitch intervals. The regularity of the glitches, i.e. almost constant magnitude and time interval between the events, in PSR J0537--6910 and the Vela entails that glitch activity described by equation (\ref{glitchac}) is approximately given by the slopes of lines in the graphs of Figure \ref{cumulative} for these pulsars. On the other hand, large coverage of glitch sizes and unequal repetition timescales for PSRs J1341-6220 and J1740-3015 suggest that different kinds of glitch mechanisms come into play at different times or a combination of two mechanisms appear with variable efficiency. If one assumes that some stellar component with moment of inertia $I_{\rm gl}$ stores angular momentum by average spin-down rate $\langle|\dot\nu|\rangle$ during long-run of observations containing many glitches, then the rate of angular momentum transferred to the crustal component with moment of inertia $I_{\rm c}$ gives an lower limit on the fractional moment of inertia of the neutron star component participated in glitches as \cite{link99b,andersson12,chamel13}
\begin{equation}
    \frac{I_{\rm gl}}{I_{\rm c}}\gtrsim\frac{\dot{\nu}_{\rm g}}{\langle|\dot\nu|\rangle}.
\label{fracI}
\end{equation}
Equation (\ref{fracI}) is applied to the entire pulsar glitch population \cite{ho15,pizzochero17,basu18,parmar22,montoli21} and gives the result of a few percent implying only crustal superfluid and probably some part of the outer core region of a neutron star are involved in glitches when crustal entrainment effect has been properly taken into consideration \cite{erbil14,montoli20,sourie20}.

The importance of continuously searching for new pulsar glitches through timing observations is self-evident. With new surveys being conducted by existing and under-construction radio telescopes around the world, the number of known pulsars is expected to increase steadily. It's definitely worth mentioning that the world's largest single-dish radio telescope, FAST, has discovered more than 600 new pulsars since it was built in 2016. In addition, monitoring programs to capture a glitch "live" and improved glitch-finding algorithms are being developed. Taking all these together, in coming years many production of increasing numbers of new glitches discoveries will be seen.

\authorcontributions{Conceptualization, S.Q.Z., E.G., J.P.Y. and M.Y.G.; methodology, S.Q.Z. and E.G.; software, S.Q.Z. and J.P.Y.;  data curation,  S.Q.Z.; writing---original draft preparation, S.Q.Z. and E.G.; writing---review and editing, S.Q.Z., E.G., J.P.Y., M.Y.G. and C.Y.; visualization, S.Q.Z.; supervision, S.Q.Z., E.G., J.P.Y., M.Y.G. and C.Y.; project administration, C.Y., J.P.Y. and M.Y.G. All authors have read and agreed to the published version of the manuscript.}

\funding{This research was funded by the National Natural Science Foundation of China (NSFC) grant via NSFC-11373064, 11521303, 11733010, 11873103, U2031121, 11873080, U1838201, U1838202, U1838104, U1938103, and U1938109. EG is partially supported by NSFC programme via 11988101.}

\conflictsofinterest{The authors declare no conflict of interest.} 

\reftitle{References}


\externalbibliography{yes}
\bibliography{biblio}
\end{document}